\documentclass{article}

\usepackage{arxiv}

\usepackage[utf8]{inputenc} 
\usepackage[T1]{fontenc}    
\usepackage{hyperref}       
\usepackage{url}            
\usepackage{booktabs}       
\usepackage{amsfonts}       
\usepackage{nicefrac}       
\usepackage{microtype}      
\usepackage{lipsum}

\usepackage{cite}
\usepackage{amsmath,amssymb,amsfonts}
\usepackage{algorithmic}
\usepackage{graphicx}
\usepackage{textcomp}
\def\BibTeX{{\rm B\kern-.05em{\sc i\kern-.025em b}\kern-.08em
    T\kern-.1667em\lower.7ex\hbox{E}\kern-.125emX}}
\usepackage{siunitx}
\usepackage{import}
\usepackage{float}
\usepackage{wrapfig}
\usepackage[font=small,labelfont=bf]{caption}
\usepackage{subcaption}
\usepackage{mathrsfs}
\usepackage{mathtools}
\usepackage{multirow}
\usepackage{lineno}
\usepackage{etoolbox}
\usepackage{setspace}

\usepackage[table]{xcolor}

\newcommand\norm[1]{\left\lVert#1\right\rVert}
\renewcommand\b[1]{\boldsymbol{\mathbf{#1}}}

\DeclareMathOperator*{\argmax}{argmax}
\DeclareMathOperator*{\argmin}{argmin}

\DeclareMathOperator{\SSIM}{SSIM}

\DeclareGraphicsExtensions{.pdf}

\newcommand{\beginsupplement}{%
        \setcounter{table}{0}
        \renewcommand{\thetable}{S\arabic{table}}%
        \setcounter{figure}{0}
        \renewcommand{\thefigure}{S\arabic{figure}}%
     }

\immediate\write18{texcount -sum=1,1,1,1,1,1,1 -merge -incbib -dir -utf8 main.tex > main.wcTotal }
\immediate\write18{texcount -sub=none -merge -incbib -dir -utf8 main.tex | grep _main_ | grep -oE '[0-9]+' | python -c "import sys; print(sum(int(l) * w for l, w in zip(sys.stdin, [1, 1, 0, 0, 0, 0, 0])))" > main.wcManuscript }
\immediate\write18{texcount -sub=none -merge -incbib -dir -utf8 main.tex | grep Abstract | grep -oE '[0-9]+' | python -c "import sys; print(sum(int(l) * w for l, w in zip(sys.stdin, [1, 1, 0, 0, 0, 0, 0])))" > main.wcAbstract }

\newcolumntype{P}[1]{>{\centering\arraybackslash}p{#1}}

\title{Reconstructing unseen modalities and pathology with an efficient Recurrent Inference Machine}

\author{
Dimitrios Karkalousos \\ 
Department of Biomedical Engineering \& Physics\\
Amsterdam UMC, University of Amsterdam\\
Meibergdreef 9, Amsterdam, Netherlands \\
\texttt{d.karkalousos@amsterdamumc.nl} \\
\And
Kai L\o nning \\
Department of Radiation Oncology\\
Netherlands Cancer Institute\\
Spinoza Centre for Neuroimaging\\
Amsterdam, Netherlands \\
\And
Hanneke E. Hulst \\
Department of Anatomy \& Neurosciences\\
Amsterdam UMC, Vrije Universiteit Amsterdam\\
MS Center Amsterdam, Amsterdam Neuroscience \\
\And
Serge O. Dumoulin \\
Department of Experimental Psychology\\
Utrecht University, Helmholtz Institute\\
Department of Experimental \& Applied Psychology\\
VU University\\
Spinoza Centre for Neuroimaging\\
Amsterdam, Netherlands \\
\And
Jan-Jakob Sonke \\
Department of Radiation Oncology\\
Netherlands Cancer Institute\\
Amsterdam, Netherlands \\
\And
Frans M. Vos \\
Department of Imaging Physics\\
Delft University of Technology\\
Department of Radiology\\
Amsterdam UMC, Academic Medical Center\\
\And
Matthan W.A. Caan \\
Department of Biomedical Engineering \& Physics\\
Amsterdam UMC, University of Amsterdam\\
Meibergdreef 9, Amsterdam, Netherlands \\
\texttt{m.w.a.caan@amsterdamumc.nl}
}

\begin{document}
\maketitle

\begin{abstract}
\normalsize
Objective: To allow efficient learning using the Recurrent Inference Machine (RIM) for image reconstruction whereas not being strictly dependent on the training data distribution so that unseen modalities and pathologies are still accurately recovered.
Methods: Theoretically, the RIM learns to solve the inverse problem of accelerated-MRI reconstruction whereas being robust to variable imaging conditions. The efficiency and generalization capabilities with different training datasets were studied, as well as recurrent network units with decreasing complexity: the Gated Recurrent Unit (GRU), the Minimal Gated Unit (MGU), and the Independently Recurrent Neural Network (IndRNN), to reduce inference times. Validation was performed against Compressed Sensing (CS) and further assessed based on data unseen during training. A pathology study was conducted by reconstructing simulated white matter lesions and prospectively undersampled data of a Multiple Sclerosis patient.
Results: Training on a single modality of 3T $T_1$-weighted brain data appeared sufficient to also reconstruct 7T $T_{2}^*$-weighted brain and 3T $T_2$-weighted knee data. The IndRNN is an efficient recurrent unit, reducing inference time by 68\% compared to CS, whereas maintaining performance. The RIM was able to reconstruct lesions unseen during training more accurately than CS when trained on $T_2$-weighted knee data. Training on $T_1$-weighted brain data and on combined data slightly enhanced the signal compared to CS.
Conclusion: The RIM is efficient when decreasing its complexity, which reduces the inference time, whereas still being able to reconstruct data and pathology that was unseen during training.
\end{abstract}


\section{Introduction}
\label{sec:introduction}
Deep Learning (DL) is a new approach to reconstructing MR images from sparsely sampled k-space data. It holds great promise for achieving shorter acquisition and faster reconstruction times. 
Essentially, deep learning approaches \emph{learn} how to solve the reconstruction problem, instead of applying a predefined sparsifying transform, as is standard in Compressed Sensing (CS) image reconstruction \cite{doi:10.1002/mrm.21391}. 
Below we will outline how the field is quickly developing on the path towards broad applicability in clinical practice.

Over the last few years, a large number of DL network architectures has been introduced.
The U-net is a traditional approach that was initially proposed for segmentation purposes \cite{DBLP:journals/corr/RonnebergerFB15}, but is now also widely used for image reconstruction \cite{aeb8599cbbab479e9fc537295391f414, 7949028}.
Alternatively, deep residual learning networks, composed of separate magnitude and phase networks, were shown to learn and resolve the aliasing artifacts resulting from the random undersampling \cite{8329428}.
To make residual networks easier to optimize and achieve higher accuracy, the layers were reformulated to learn residual functions with reference to the layer inputs, instead of learning unreferenced functions \cite{DBLP:journals/corr/HeZRS15, WU202093}.
Another paper applied a neural network trained by the Alternating Direction Method of Multipliers (ADMM) algorithm in order to optimize a general CS-MRI model using DL for parameter learning \cite{DBLP:journals/corr/YangSLX17}.
The Deep Cascade of CNNs approach stacked Convolutional Neural Networks (CNN) together for reconstruction, applying scattered k-space correction layers designed to maintain data consistency \cite{DBLP:journals/corr/SchlemperCHPR17a}.
The Convolutional Recurrent Neural Network (CRNN) learns temporal dependencies and embeds the structure of traditional iterative algorithms with a small number of parameters, to achieve faster inference times and higher accuracy \cite{Qin2019}. 
Other approaches use unrolling-based DL methods aiming to solve the inverse problem by learning a set of parameters associated with a prior model with filter kernels, activation functions, and data term weights \cite{doi:10.1002/mrm.26977, Adler, 10.1007/978-3-030-32248-9_3, Hammernik2019}. The use of Generative Adversarial Networks (GANs) for reconstruction was also advocated in order to remove aliasing artifacts \cite{8417964,8233175,DBLP:journals/corr/abs-1805-10704}, but generalization capabilities have not yet been demonstrated for GANs. 
In order to accelerate parallel imaging, a MultiLayer Perceptron (MLP) has been built, exploiting information from multiple receiver coils with different spatial sensitivities \cite{10.1371/journal.pone.0189369}. Applying domain-transform manifold learning, Zhu et al. \cite{Zhu2018} proposed a scheme called AUtomated TransfOrm by Manifold APproximation (AUTOMAP) that directly maps the sparsely sampled measurements to the fully sampled image. 
To reconstruct multi-coil multi-contrast data, Joint Virtual Coils were introduced, to encode intensity and phase variations across images \cite{doi:10.1002/mrm.27076}. 
Yet another paper trained CNNs on AutoCalibration Signal (ACS) data for k-space interpolation, in order to reconstruct both 2D and 3D acquisitions with uniform undersampling \cite{doi:10.1002/mrm.27420}.
Also, a Model Based Deep Learning (MoDL) framework was proposed which gains efficiency through an inverse problem solver consisting of a CNN-based regularization prior, sharing weights across iterations \cite{DBLP:journals/corr/abs-1712-02862}.
The importance of the acquisition protocol as deep prior was recently explored in \cite{Liu2019} and further addressed in \cite{Zhou2020DuDoRNetLA} using a Dual-Domain Recurrent Network, with $T_1$-weighted images as deep prior in both image domain and k-space domain. Cross-domain learning has also been proposed in \cite{Wang2019} using a multi-supervised loss to combine sufficient information of the frequency and the spatial domain.

Whereas the methods mentioned above adequately solve the problem of image reconstruction in some situations, they have not been shown to always operate robustly under variable imaging conditions, with multiple imaging sequences and in the presence of pathological conditions. Initial work was done using the Variational Network, demonstrating that it generalizes to different contrasts and sampling in knee imaging \cite{doi:10.1002/mrm.27355}.
An end-to-end method was recently proposed to extend the Variational Network by learning sensitivity maps as part of the reconstruction process \cite{Sriram2020EndtoEndVN}. Multi-channel end-to-end reconstruction has also been proposed in \cite{Wang2020}, using complex convolutions to exploit correlations between the real and imaginary part of MR images and identify the relationship between undersampled and fully sampled data.
Furthermore, Liu et al. proposed a framework to optimize CS-MRI and address its robustness against noise observed in actual MR-imaging \cite{Liu2019}. 
A relevant issue in relation to DL image reconstruction is that fast inference times are needed whereas maintaining accurate performance. A few approaches have been proposed balancing these aspects, including domain adaption from CT and synthetic MRI for radially sampled data \cite{doi:10.1002/mrm.27106}. 
Other approaches are trying to reconstruct images without access to ground-truth data, using unsupervised learning for an unrolled optimization \cite{Tamir} or using Variational Auto-Encoders \cite{Tezcan2019}.

In this work we present major advances of our previously proposed Recurrent Inference Machine (RIM) in reconstructing heterogeneous MRI data \cite{LONNING201964}. We aim to extend our method by exploring fast inference times and the generalization capabilities of a small and efficient model. 
To this end, we propose and evaluate the efficiency of different recurrent units and also test the importance of the choice of the loss function \cite{cc2825e6939c4d09981c5d9c7ce29823, Zhao2015LossFF}.
Furthermore, we will test the model's accuracy and dependency in handling unseen pathology during training. We will conduct experiments on lesions in the brain, using synthetic data, and images of a Multiple Sclerosis patient. 

Extensive comparison of the RIM's performance against state-of-the-art methods has been reported and verified elsewhere, and is out of the scope of this work.
More specifically, L\o nning et al. \cite{LONNING201964} show a comprehensive evaluation of the RIM in various imaging conditions and multiple acceleration factors against the U-Net and CS. In \cite{Putzky2019InvertTL} the RIM was evaluated against the U-Net and a very deep variation of our model the invertible RIM \cite{Putzky2019iRIMAT}, using 1D cartesian undersampling masks on singlecoil and multicoil data within the context of the fastMRI challenge \cite{doi:10.1002/mrm.28338}.
Circumstantial comparison of the invertible RIM (being one the three winning solutions \cite{Wang2019PyramidCR, Pezzotti2020AnAI}) against several methods can be found in the leaderboard and the results of the challenge. 
Finally, in the recent Multi-channel MR Image Reconstruction Challenge \cite{SOUZA2018482} the RIM was the winning solution, compared against quite a few dual-domain approaches \cite{Souza, Souza2019, Souza2019SIBGRAPI} and variations of them. Results about the RIM's performance and implementation can be found in our public repository \cite{DIRECTTOOLKIT}. 

\section{Methods}
\subsection{Accelerated MRI Reconstruction}
\begin{figure*}
\center
\centerline{\includegraphics[width=\textwidth]{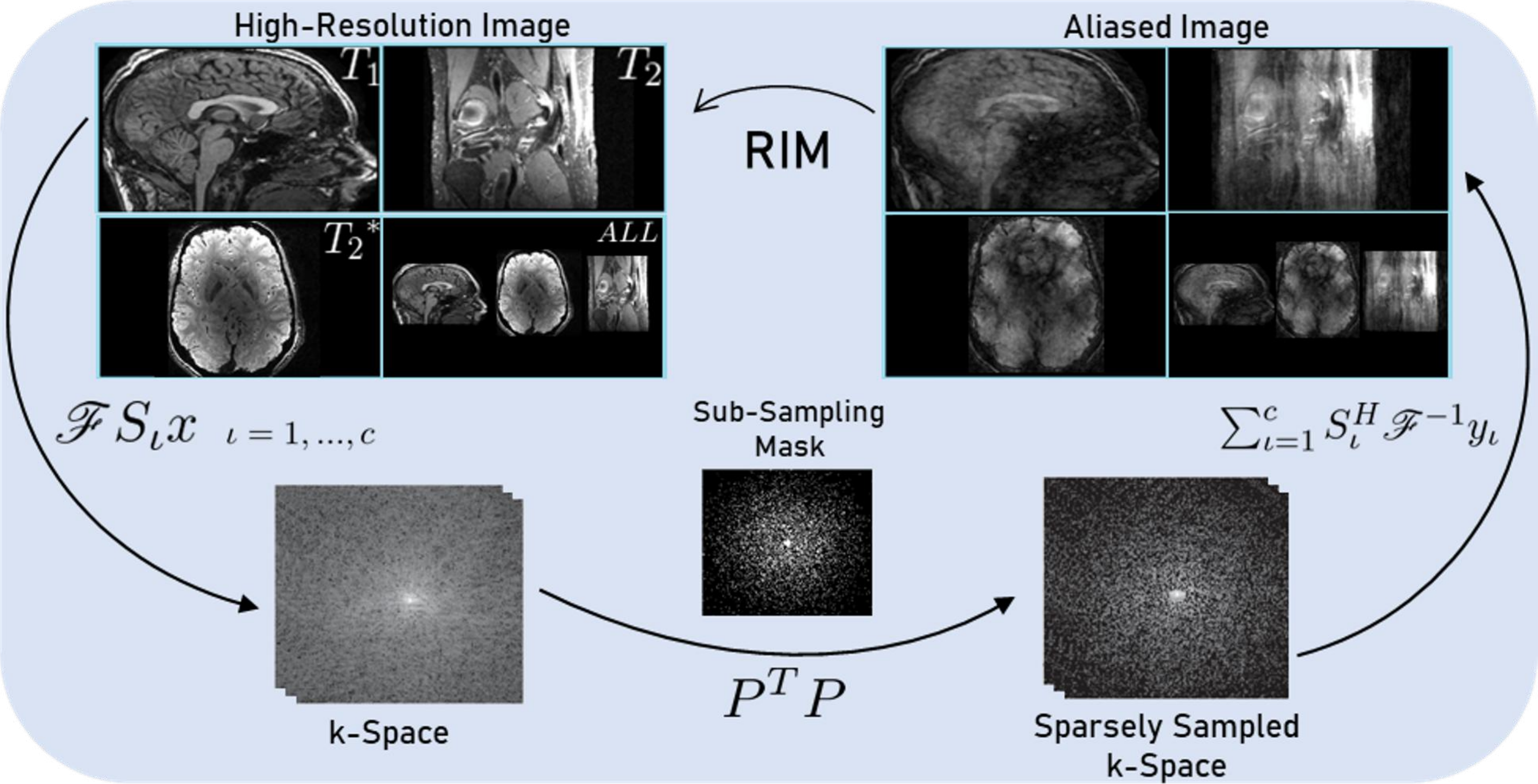}}
\caption{The goal in accelerated-MRI reconstruction is to find the inverse transform for recovering a high-resolution image (top-left) from the sparsely sampled measurements (bottom-right). A forward model starts from the high-resolution image representation, which is Fourier transformed into k-space (bottom-left), and subsequently sub-sampled (bottom-middle) to obtain sparsely sampled measurements (bottom-right). Through the inverse Fourier transform an aliased image is obtained (top-right). The RIM solves the inverse problem over multiple time-steps starting from the aliased image as initial estimate.}
\label{fig:mri_measurement}
\end{figure*}

As illustrated in Figure~\ref{fig:mri_measurement}, the goal in accelerated-MRI reconstruction is to find an inverse transform for mapping sparsely sampled k-space measurements (bottom-right) to a high resolution image (top-left). Let the set of the fully sampled data be denoted as $\mathbf{x}\in{\mathbb{C}^n}$, and let $\mathbf{y_{\iota}}\in{\mathbb{C}^m}$, $m\ll{n}$, be the set of the sparsely sampled data in k-space of the $\it{\iota}$-th coil, from $\mathit{c}$ receiver coils. Effectively, the sparse sampling facilitates reduced the scan time. 

A forward model describes how the measured data are obtained from a underlying reference image. Essentially, by applying the Fourier transform $\mathscr{F}$, the reference image $\mathbf{x}$, is projected on the k-space where a sub-sampling mask $\mathbf{P}$, discards some fraction of values to model the sparse sampling,
\begin{equation}\label{eq:corruption_process} \mathbf{y}_{\it{\iota}}= \mathbf{P} \mathscr{F} \mathbf{S}_{\it{\iota}}^{H} \mathbf{x}+\mathbf{n}_{\it{\iota}}, \hspace{1em} \it{\iota}=1,...,\mathit{c}, \end{equation}
where $\mathbf{S}_{\it{\iota}}$ is the sensitivity map, a diagonal matrix of the spatial sensitivity that scales every pixel of the reference image $\mathbf{x}$ by a complex number and $\mathbf{H}$ is the adjoint operator. 
The signals are assumed to be distorted by additive, normally distributed noise $\mathbf{n}_{\it{\iota}} \sim N\left(0, I{\sigma}^2\right) + iN\left(0, I{\sigma}^2\right)$, $i^2=-1$, resulting from further unspecified, accumulated measurement errors. 
\hfill \break
The inverse transform of restoring the high-resolution reference image from the set of the sparsely sampled measurements, can be found through the Maximum A Posteriori (MAP) estimator, given by 
\begin{equation}\label{eq:map}\mathbf{x}_{MAP} = \argmax_{\mathbf{x}}\left\{\log p\left({\mathbf{y}}\mid{\mathbf{x}}\right) + \log p\left({\mathbf{x}}\right) \right\},\end{equation}
which is the maximization of the sum of the log-likelihood and the log-prior distribution of $\mathbf{y}$ and $\mathbf{x}$, respectively. Here, the log-prior distribution represents an MR-image’s likely appearance which acts as a regularization term. Conventionally, Eq. \ref{eq:map} is reformulated as the following optimization problem,
\begin{equation}\label{eq:regularized_map}
\mathbf{x}_{MAP} =
\argmin_{\mathbf{x}}\left\{\sum_{\it{\iota}=1}^{\mathit{c}} \mathbf{d}({\mathbf{y}_{\it{\iota}}},{\mathbf{P} \mathscr{F}\mathbf{S}_{\it{\iota}}^{H}\mathbf{x}}) + \b{\lambda} \mathbf{R}(\mathbf{x})\right\},
\end{equation}
where $\mathbf{d}$ evaluates the data consistency between the reconstruction and the measurements, and $\mathbf{R}$ is a regularizer preventing overfitting. The regularization factor $\b{\lambda}$ limits the solution space and incorporates prior knowledge.
\hfill \break
The log-likelihood distribution in Eq. \ref{eq:map} reflects the error distribution between the reconstructed image and the acquired measurements. 
It can be seen as a data fidelity term, corresponding to the data consistency in Eq. \ref{eq:regularized_map}.
Assuming Gaussian distributed data (see above), this term is given by:
\begin{equation}\label{eq:log_likelihood_corr_d}\log p\left({\b y}\mid{\b x}\right) = \frac{1}{{\sigma}^2} \sum_{\it{\iota}=1}^{\mathit{c}} {\parallel {\mathbf{P} \mathscr{F} \mathbf{S}_{\it{\iota}}^{H} \mathbf{x}} - \mathbf{y}_{\it{\iota}} \parallel}_{2}^{2}.\end{equation}

\subsection{Recurrent Inference Machine (RIM)}
\begin{figure*}
\captionsetup[subfigure]{justification=centering}
    \centering
    \begin{subfigure}[]{0.3\textwidth}
        \centering
        \centerline{\includegraphics[width=\columnwidth]{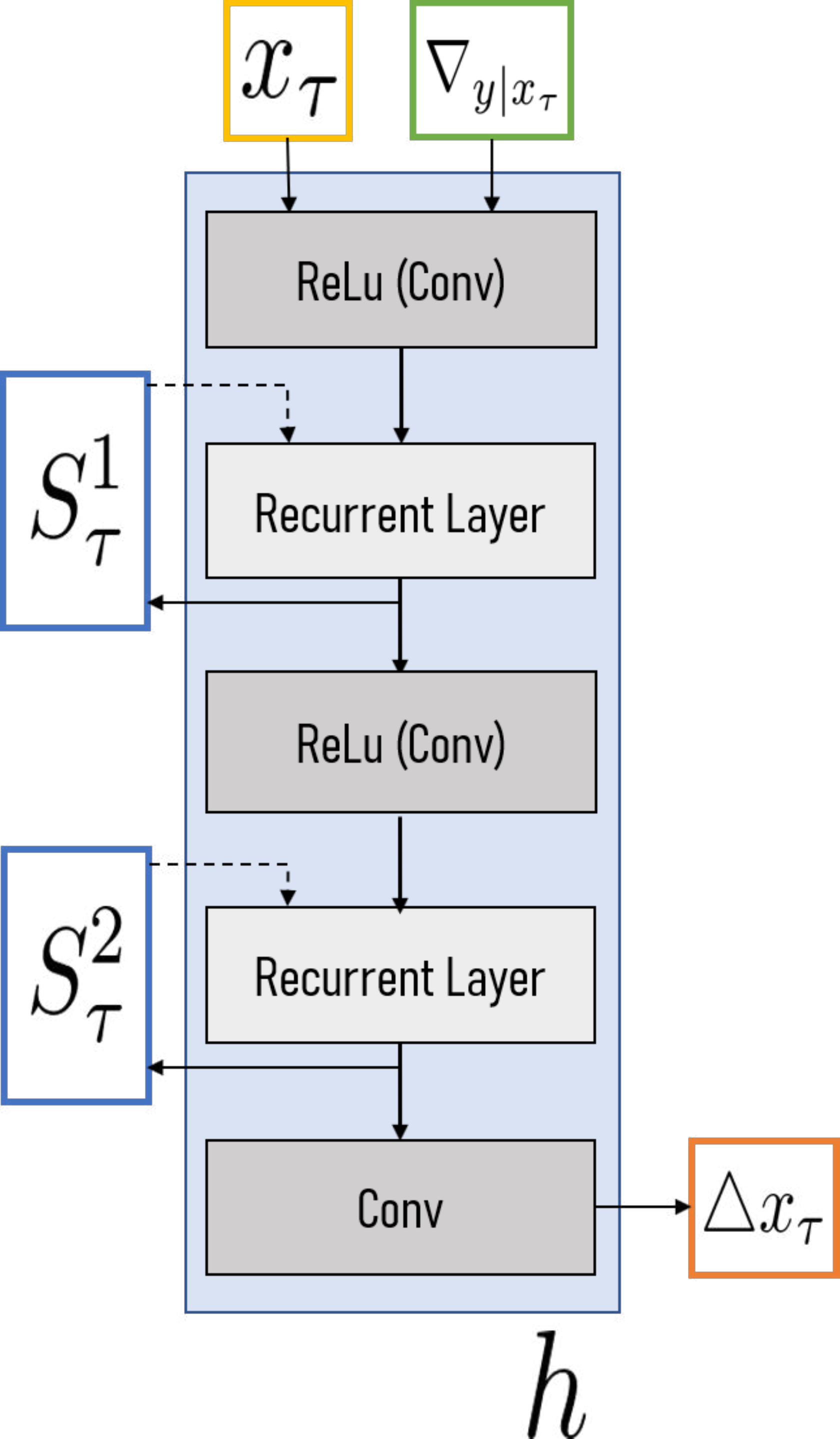}}
        \caption{RIM}
        \label{fig:rim}
    \end{subfigure}
    ~
    \centering
    \begin{subfigure}[]{0.6\textwidth}
        \centering
        \centerline{\includegraphics[width=\columnwidth]{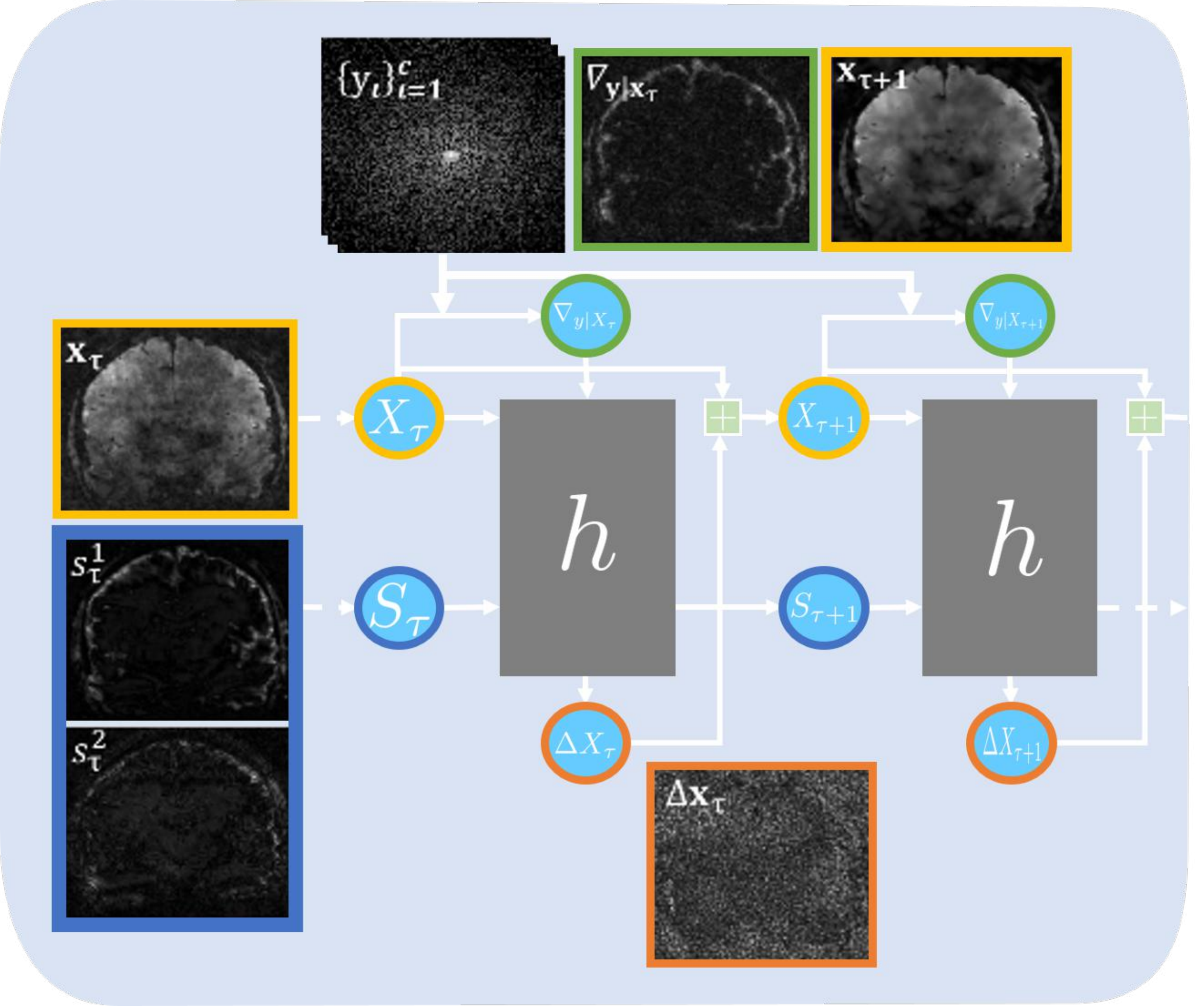}}
        \caption{Unfolded RIM}
    \end{subfigure}
    \\
    \vspace{1em}
    \centering
    \begin{subfigure}[]{\textwidth}
        \centering
        \centerline{\includegraphics[width=\columnwidth]{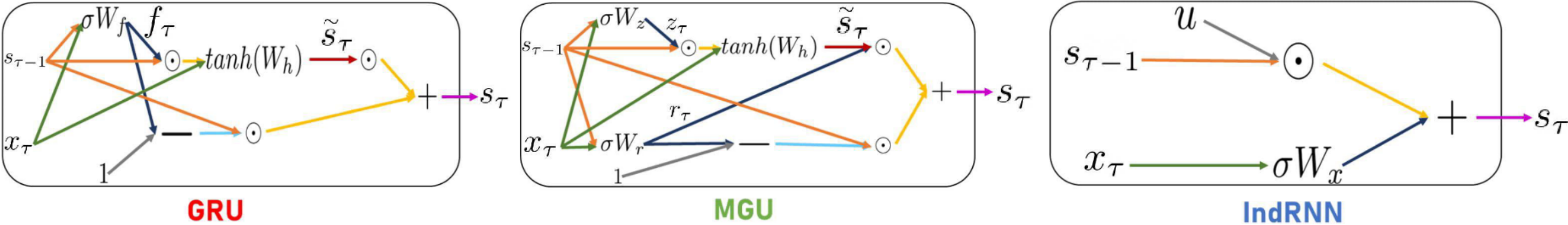}}
        \caption{Computational graphs of the GRU, the MGU, and the IndRNN}
        \label{fig:GRU_MGU_IndRNN}
    \end{subfigure}    
    \caption{Architecture of the Recurrent Inference Machine (RIM). (a) The RIM uses an alternating sequence of convolutional and recurrent layers. Updates are passed to the next time-step through the dotted lines, whereas the bold lines represent the connections within one time-step. (b) Unfolded update functions of the RIM for two iterations. The RIM takes as input the concatenation of the the gradient of the log-likelihood and the current reconstruction and outputs the complex-valued image update $\Delta{\mathbf{x}_{\it{\tau}}}$. (c) Each computational graph reflects the complexity of the unit for the recurrent layers of the RIM. Shown are the Gated Recurrent Unit (GRU), the Minimal Gated Unit (MGU), and the Independently Recurrent Neural Network (IndRNN).
    } 
\end{figure*}

The Recurrent Inference Machine (RIM) was first proposed as general inverse problem solver \cite{DBLP:journals/corr/PutzkyW17}. 
Particularly, the RIM uses a sequence of alternating convolutional and recurrent layers (Fig. \ref{fig:rim}). 
The Gated Recurrent Unit (GRU) \cite{DBLP:journals/corr/ChoMGBSB14}, the Minimal Gated Unit (MGU) \cite{DBLP:journals/corr/ZhouWZZ16}, and the Independently Recurrent Neural Network (IndRNN) \cite{DBLP:journals/corr/abs-1803-04831}, that will be introduced in more detail below, each represent optional configurations for the recurrent layers (Fig. \ref{fig:GRU_MGU_IndRNN}). 

The RIM targets to optimize the regularised problem (Eq. \ref{eq:regularized_map}) over $\mathit{t}$ time-steps . At each time-step $\it{\tau}$ the gradient of the log-likelihood (Eq. \ref{eq:log_likelihood_corr_d}) and the current reconstruction are concatenated and given as input to the RIM, which yields an incremental step $\Delta{\mathbf{x}_{\it{\tau}}}$ to take as output in order to improve the current reconstruction state.
The log-likelihood gradient at time-step $\it{\tau}$ is given by:
\begin{equation}\label{log-likelihood-grad}
\nabla_{\mathbf{y}\vert\mathbf{x}_{\it{\tau}}}\coloneqq\frac{1}{\sigma^2}\sum_{\it{\iota}=1}^{\mathit{c}} \mathbf{S}_{\it{\iota}}^{H} \mathscr{F}^{-1} \mathbf{P}^{\mathrm{T}}\left(\mathbf{P}\mathscr{F} \mathbf{S}_{\it{\iota}} \mathbf{x}_{\it{\tau}}-\mathbf{y}_{\it{\iota}}\right).\end{equation}
Notice that the network merely evaluates the gradient of the log-prior distribution by passing the current estimate $\mathbf{x}_{\it{\tau}}$ as an input to the network.
By doing so, the prior itself is never explicitly evaluated \cite{DBLP:journals/corr/PutzkyW17}.

Let the RIM's update function be denoted by $\mathbf{h}$, then the RIM's update equations are given by
\begin{equation}
\label{eq:update}
    \begin{aligned}
        \mathbf{s_0}&=\mathbf{0},&~
        \mathbf{x_0}&=\sum_{\it{\iota}=1}^{\mathit{c}}\mathbf{S}_{\it{\iota}}^{H}\mathscr{F}^{-1}\mathbf{P}^{\mathrm{T}}\mathbf{y}_{\it{\iota}},\\
        \mathbf{s}_{\it{\tau}+1}&=\mathbf{g}\left(\nabla_{\mathbf{y}\vert\mathbf{x}_{\it{\tau}}},\mathbf{x}_{\it{\tau}},\mathbf{s}_{\it{\tau}}\right),&~
        \mathbf{x}_{\it{\tau}+1}&=\mathbf{x}_{\it{\tau}}+\mathbf{h}\left(\nabla_{\mathbf{y}\vert\mathbf{x}_{\it{\tau}}},\mathbf{x}_{\it{\tau}},\mathbf{s}_{\it{\tau}+1}\right),
    \end{aligned}
\end{equation}
for $0\leq{\it{\tau}}<\mathit{t}$. 
$\mathbf{s}_{\it{\tau}}$ and $\mathbf{x}_{\it{\tau}}$ represent the internal and external states of the network respectively at time-step $\it{\tau}$, whereas $\mathbf{g}$ is a function representing the part of the network that creates the next internal state $\mathbf{s}_{\it{\tau}+1} = \left(\mathbf{s}^1_{\it{\tau}}, \mathbf{s}^2_{\it{\tau}}\right) $ during iterations. 

The internal states act as in conventional RNNs, making the activation at each time-step dependent on that of the previous time. As such the internal states make the model more input independent and robust against overfitting.
Observe that the initial estimate $\mathbf{x_0}$ is the inverse Fourier transform of the sparsely sampled measurements (Fig.\ref{fig:mri_measurement} top-right) and $\mathbf{s_0}$ is initialized to zero \cite{LONNING201964}. 

The first two convolutional layers of the RIM are being activated by the ReLU function, whereas the recurrent layers maintain the internal states. The first convolutional layer has a kernel size of 5$\times$5 whereas the next two convolutional layers have kernel sizes of 3$\times$3. All inputs are zero-padded to maintain image sizes. The final convolutional layer outputs a complex-valued image update $\Delta{\mathbf{x}_{\it{\tau}}}$. 

The hyper-parameter $\mathit{F}$ denotes the number of features extracted from the recurrent layers’ internal states and the convolutional layers, and it is kept constant throughout the RIM. The recurrent layers' weights are shared across image pixels, whereas they differ across the convolutional layers' feature maps, enabling reconstructing images of arbitrary size.

\subsection{Recurrent Units}
\label{subsec:rec_layers}
The recurrent nature of the RIM allows the network to have a notion of memory, allowing it to discard information deemed not useful and on the other hand retain information that is considered useful.
Multiple units were previously proposed as candidates for the recurrent layers, with varying complexity and numbers of parameters.
In this work, we implemented three different types of these units, aiming to minimize the number of model parameters whereas maintaining sufficient complexity for accurate reconstruction. 
These recurrent units are graphically depicted in Fig. \ref{fig:GRU_MGU_IndRNN}.

First, the GRU \cite{DBLP:journals/corr/ChoMGBSB14} has two gating units, the reset gate and the update gate. These gates control how the information flows in the network.
The update gate regulates the update to a new hidden state, whereas the reset gate controls the amount of information to forget. Both gates act in a probabilistic manner. 
Second, the MGU \cite{DBLP:journals/corr/ZhouWZZ16} is a simplified version of the GRU, having the smallest possible number of gates in any gated unit. 
To this end, a single coupled gate is defined, named the forget (or update) gate.
The coupling of gates may lead to reduced variance in the activations or backpropagated gradients of the network.
Therefore, using the Xavier weight initialization \cite{pmlr-v9-glorot10a} can ameliorate this potential problem.
Third, the IndRNN \cite{DBLP:journals/corr/abs-1803-04831} is a shallow network designed to address gradient decay over iterations, following an independent neuron connectivity within a recurrent layer. 

Separate RIMs were built with each of these recurrent units. Henceforth, we will refer to these as GRIM, MRIM, and IRIM, as the GRU, the MGU, and the IndRNN are used, respectively.

\subsection{Loss Function}
\label{subsec:loss_function}
In our experiments we applied two loss functions. The first one was based on the $\ell_2$-norm, also known as Mean Squared Error (MSE), which represents the average squared difference between the estimate $\mathbf{x}_{\it{\tau}}$ and the reference image $\mathbf{x}$. Specifically, we used the weighted sum of the MSE averaged over all iterations to yield the total $\ell_2$-loss, given by
\begin{equation}\label{eq:l2_loss}\mathcal{L}^{\ell_2}\left(\mathbf{x}_{\mathit{t}}\right) = \frac{1}{\mathit{n}\mathit{t}}\sum_{\it{\tau}=1}^{\mathit{t}}\mathbf{w}_{\it{\tau}}\norm{\mathbf{x}_{\it{\tau}} - \mathbf{x}}_2^2,\end{equation}
where $\mathit{n}$ is the total number of pixels of the image and $\mathbf{w}_{\it{\tau}}$ is a weight vector of length $\mathit{t}$ determining the importance attributed to the loss at time-step $\it{\tau}$. The weights were calculated by setting $\mathbf{w}_{\it{\tau}}=10^{-\frac{\mathit{t}-{\it{\tau}}}{\mathit{t}-1}}$ to prioritise the latter time-steps during reconstruction. 

Whereas the $\ell_2$-norm is an efficient metric for low dimensional problems, it may not give meaningful results for a high-dimensional problem such as image reconstruction \cite{10.1007/3-540-44503-X_27}. Particularly, the probability of all dimensions being equally relevant conventionally decreases with increasing data dimensionality. More formally stated, for all $\mathcal{L}_\mathit{k}$-norms with $\mathit{k} \geq 1$, the ratio of the variance of the length of any point vector to the mean point vector converges to zero \cite{doi:10.1002/sam.11161} with increasing dimensions, which might lead to gradients vanishing or exploding \cite{DBLP:journals/corr/abs-1211-5063}.

Therefore, we also trained models using the $\ell_1$-norm as loss function, also known as Least Absolute Deviations. The $\ell_1$-norm represents the sum of the absolute difference $|\cdot|$ between the estimate $\mathbf{x}_{\it{\tau}}$ and the reference image $\mathbf{x}$:
\begin{equation}\label{eq:l1_loss}\mathcal{L}^{\ell_1}\left(\mathbf{x}_{\mathit{t}}\right) = \frac{1}{\mathit{n}\mathit{t}}\sum_{\it{\tau}=1}^{\mathit{t}}\mathbf{w}_{\it{\tau}}\mid{\mathbf{x}_{\it{\tau}} - \mathbf{x}}\mid.\end{equation}

\subsection{Datasets}
\label{subsec:Datasets}
\begin{table*}[]
    \centering
    \caption{Scan parameters for different experiments. Target anatomy, contrast, scan and field strength are given, with resolution (res), Field-of-View (FOV), time in minutes (with acceleration factor), number of coils (ncoils) and other scan parameters.}
    \begin{tabular}{lllccl}
        \cmidrule(l){1-6}
        \textbf{scan, field strength} & \textbf{res (mm)} & \textbf{FOV (mm)} & \textbf{time (acc)} & \textbf{ncoils} & \textbf{parameters} \\
        \cmidrule(l){1-6}
        \multicolumn{6}{l}{\cellcolor{lightgray}Training, validation}\\
        brain 3D $T_1$ MPRAGE, 3T & 1.0x1.0x1.0 & 256x256x240 & 10.8 & 32 & FA \ang{9}, TFE-factor 150, TI=900ms \\
        brain 3D $T_2^*$ FLASH, 7T & 0.7x0.7x0.7 & 224x224x180 & 22.5 & 32 & FA \ang{12}, TR=23.4ms, TE=3-21ms, 6 echoes \\
        knee 3D $T_2$ TSE, 3T & 0.5x0.5x0.6 & 160x160x154 & 15.3 & 8 & FA \ang{90}, TR=1550ms, TE=25ms \\
        \cmidrule(l){1-6} 
        \multicolumn{6}{l}{\cellcolor{lightgray}Evaluation} \\
        brain 3D $T_2$ TSE, 3T & 1.0x1.0x1.0 & 250x250x180 & 1.9 & 32 & TR=3000ms, TE=280 ms, TSE-factor 110 \\
        \cmidrule(l){1-6} 
        \multicolumn{6}{l}{\cellcolor{lightgray}Pathology study} \\
        brain 3D FLAIR, 3T & 1.0x1.0x1.1 & 224x224x190 & 5.0 (6x) & 32 & TR=4800ms, TE=350ms, TI=1650ms \\
        \bottomrule
    \end{tabular}
    \label{table:scan_parameters}
\end{table*}

For our experiments, we created multiple datasets. 
In summary: (1) for training, validation, and evaluation, fully sampled data were acquired and retrospectively undersampled; (2) for independent evaluation, data with a different contrast were acquired and retrospectively undersampled; (3) to study the effect on small, previously unseen lesions, 3D FLAIR scans were obtained with prospective undersampling from patients suffering from Multiple Sclerosis (MS). Scanning parameters are given in \ref{table:scan_parameters}. 

For objective (1) three fully sampled raw complex-valued multi-coil datasets were obtained. The first two of these datasets were acquired in-house. To this end, 12 healthy subjects were included, from whom written informed consent (under an institutionally approved protocol) was obtained beforehand. The ethics board of Amsterdam UMC declared that this study was exempt from IRB approval. All 12 subjects were imaged on two scanners as follows.
3D $T_1$-weighted brain imaging was performed on a 3.0T Philips Ingenia scanner (Philips Healthcare, Best, The Netherlands) in Amsterdam UMC. Additionally, 3D $T_{2}^*$-weighted brain imaging was done on a 7.0T Philips Achieva scanner (Achieva, Philips Healthcare, Cleveland, USA) also in Amsterdam UMC.
These data were visually checked to ascertain that they were not affected by motion artifacts, after which raw data were exported and stored for offline reconstruction experiments.
For those two datasets the training sets were composed of 10 subjects, one subject was used for validation and one subject for evaluating performance depending on the training distribution. 
The third dataset was composed of 3D knee scans of 20 consecutive subjects \cite{Epperson2013}. From these data, one subject was discarded because motion artifacts were observed. The training set consisted of 17 subjects, the validation set of one subject and one subject was also used for evaluation. A combined dataset was created by combining all modalities with equal proportions using a weighted data sampler.

For independent evaluation, i.e. objective (2), a fully sampled scan was acquired on a 3.0T Philips Ingenia scanner (Philips Healthcare, Best, The Netherlands) in Amsterdam UMC of one healthy subject. As above, written informed consent (under the institutionally approved protocol) was given beforehand, also exempt from IRB approval. 
A $T_2$-weighted TSE sequence was used to acquire a brain image with a fully sampled 3D-acquisition and a resolution of \SI{1.0}{\milli\meter}. 

All datasets were Fourier transformed along the frequency encoding axis, and subsequently used as separate slices along the two phase encoding axes, along which data were retrospectively undersampled in 2D as follows.
K-space points were sampled from a two-dimensional Gaussian distribution with a Full Width at Half Maximum (FWHM) of 0.7 relative to the k-space dimensions, thereby prioritising the sampling of low frequencies, whereas creating incoherent noise due to the randomness. In that way we abide to the CS requirement of processing incoherently sampled data \cite{doi:10.1002/mrm.21391}. 
For autocalibration purposes, data points near the k-space center were fully sampled within an ellipse of which the half-axes were set to 2\% of the fully sampled region. We used acceleration factors of 4x, 6x, 8x, and 10x, and created different masks for training and evaluation, similar to related work \cite{doi:10.1002/mrm.26977, DBLP:journals/corr/abs-1709-00753, DBLP:journals/corr/YangSLX17, aeb8599cbbab479e9fc537295391f414, DBLP:journals/corr/SchlemperCHPR17a, LONNING201964}. 

Training, validation, and evaluation datasets were generated by randomly dividing slices of all subjects. The optimization of the hyperparameters, i.e. patch size, number of features and time steps, is described elsewhere \cite{LONNING201964}. The estimation of the coil sensitivities was done by using an autocalibration procedure \cite{doi:10.1002/mrm.24751}. For all datasets and each subject, the volumes were normalized with respect to the maximum magnitude. 

For every scan the Signal-to-Noise Ratio (SNR) was calculated as follows. 
The signal level was obtained by taking the mean value after thresholding the magnitude image to discard the background.
The noise level was computed as the median magnitude value within a square region of size X in the periphery of k-space, which was assumed to be dominated by imaging noise.
The ratio of these two was taken as SNR-estimates, across which the mean and standard deviation per training modality were computed.

Finally, to study the model's ability to reconstruct unseen pathology, c.f. objective (3), FLAIR data of a Multiple Sclerosis patient with known white matter lesions were used. Data were acquired on a 3.0T Philips Ingenia scanner (Philips Healthcare, Best, The Netherlands) in Amsterdam UMC within the scope of a larger, ongoing study. This study was approved by the local ethics review board and the concerned patient provided informed consent prior to imaging.
Image acquisition was six times prospectively accelerated using a pseudo-random undersampling pattern as implemented by the vendor, with a fully sampled center of 21 $\times$ 21 lines for autocalibration.

\subsection{Evaluation metrics}
For quantitative evaluation we compared normalized magnitude images derived from the complex-valued estimations $\mathbf{x}_{\it{\tau}}$ against the fully-sampled reference $\mathbf{x}$. We therefore calculated the Structural Similarity Index Measure (SSIM) \cite{1284395}, $\SSIM\left(\mathbf{x}_{\it{\tau}},\mathbf{x}\right)=\frac{\left(2\bar{\mathbf{x}_{\it{\tau}}}\bar{\mathbf{x}}+\mathit{C}_1\right)+\left(2\it{\sigma}_{\mathbf{x}_{\it{\tau}} \mathbf{x}}+\mathit{C}_2\right)}{\left(\bar{\mathbf{x}_{\it{\tau}}}^2+\bar{\mathbf{x}}^2+\mathit{C}_1\right)\left(\it{\sigma}_{\mathbf{x}_{\it{\tau}}}^2+\it{\sigma}_\mathbf{x}^2+\mathit{C}_2\right)}$, where $\bar{\mathbf{x}_{\it{\tau}}}$, $\it{\sigma}_{\mathbf{x}_{\it{\tau}}}^2$, $\bar{\mathbf{x}}$, $\it{\sigma}_\mathbf{x}^2$, denote the average and the variance of the estimation and the reference image respectively, and $\it{\sigma}_{\mathbf{x}_{\it{\tau}} x}$ the covariance. Here, $\mathit{C}_1=\left(0.01\mathit{L}\right)^{2}$ and $\mathit{C}_2=\left(0.03\mathit{L}\right)^{2}$ are regularization parameters to avoid  division by zero, whereas $\mathit{L}=\left(2^{bits}-1\right)$ represents the dynamic range of the pixel-values. 
Also, we calculated the Peak Signal-to-noise ratio (PSNR) across the images.

\subsection{Learning Efficiency }
\label{subsec:models}
The efficiency of our models was first tested by comparing inference times as a function of the size of the network. 
For a number of  time-steps $\mathit{t}$ (6, 8, 10, 12, 14, and 16) and for a number of features $\mathit{F}$ (16, 32, 64, 128, and 256), we built GRIMs, MRIMs, and IRIMs as defined in subsection \ref{subsec:rec_layers}. 
As such we assessed the inference times of reconstructing a single slice from the $T_2$-weighted TSE brain dataset and repeating this 300 times at each setting, and then we took the mean computation time over all runs. 
For fair comparison, we tested the inference time of CS using the GPU and not only the CPU as used conventionally. 

Furthermore, we tested the efficiency of our models with the two loss functions and varying types of training data. We chose to train the models with $\mathit{F}$=64 features and $\mathit{t}$=8 time-steps, applying the ADAM optimizer \cite{Kingma2015}, and set the data type to float32. Each model was trained separately on the $T_1$-weighted brain dataset, the $T_{2}^*$-weighted brain dataset, and the $T_2$-weighted knee dataset and on ALL datasets combined (ALL). Whereas doing so, data augmentation was applied by randomly cropping, rotating, flipping and mirroring patches of size 190$\times$190. For each dataset we trained each model two times, using the $\ell_2$-norm (Eq.\ref{eq:l2_loss}), and the $\ell_1$-norm (Eq.\ref{eq:l1_loss}) as loss-function, respectively. We applied the same numbers of training data across the four datasets by applying random samplings. As such, the sizes of the training sets were the same.
The performance of the trained models and the performance of CS was evaluated on unseen samples from the evaluation dataset and on the $T_2$-weighted TSE brain dataset (which was not included in the training).

CS reconstructions were performed with the BART toolbox \cite{Ong2015}. Here we used  parallel-imaging compressed-sensing reconstruction with a $\ell_1$-wavelet sparsity transform, the regularisation parameters set to $\mathit{a} = 0.005$, which was heuristically determined as a trade-off between aliasing noise and blurring. The max number of iterations set to 60.

\subsection{Robustness}
The robustness of the RIM was further assessed by reconstructing images with pathology unseen during training. To that end, a lesion simulation experiment was conducted, emulating the appearance of white matter lesions in the human brain. Additionally, an experiment was done using the prospective data from a patient with Multiple Sclerosis. Below we detail the setup of these experiments. 

\subsubsection{Simulating white matter lesions} 
White matter lesions appear as hyperintensities in a 3D-FLAIR scan. Accordingly, a single slice out of the reconstructed 3D-FLAIR scan acquired at 1mm resolution of a healthy volunteer served as the base data for the experiment.
We modelled lesions as having increased signal magnitude but negligible phase change on account of the applied fast spin-echo readout.
Lesions were generated by summing bandwidth-limited Gaussian shaped objects with $\it{\sigma}=1$ voxel, multiplied with a varying factor, to the original image intensities. This multiplication factor allowed us to study the conspicuity of lesions with decreasing image contrast.
\hfill \break
The lesions were inserted in a homogeneous area of deep white matter at one manually selected position. The maximum lesion intensity was varied from one to two times the mean intensity of the surrounding voxels. 
After inserting the lesion in the image, multi-coil data were synthesised through the forward model (Eq. \ref{eq:corruption_process}) using pre-computed coil sensitivity maps from the validation dataset (subsection \ref{subsec:Datasets}). Subsequently, complex-valued Gaussian noise was added to each coil image with a standard deviation of 5\% of the mean of the ground truth magnitude image. 
The resulting images were Fourier transformed and 4 to 8 times undersampled using 10 variable density random undersampling masks.
The resulting data were reconstructed using the best performing model (subsection \ref{subsec:models}) and compared to CS reconstruction.

\subsubsection{Multiple Sclerosis Patient Data}
The prospectively undersampled 3D-FLAIR data of a Multiple Sclerosis patient was reconstructed using the most efficient model, and CS serving as reference, given its availability as a clinical product. 
The resulting images were visually compared regarding image contrast between white matter lesions and surrounding tissue.

\section{Results}
\subsection{Learning efficiency }
In the first set of our experiments we tested the efficiency of the models on a Nvidia Tesla K40m GPU-card. \ref{table:rim_mrim_indrnn_parameters} shows an overview of the total number of model parameters as a function of the number of features (subsection \ref{subsec:models}). The IRIM has a less complex composition compared to the GRIM and the MRIM (c.f. Fig. \ref{fig:GRU_MGU_IndRNN}), such that its number of parameters is lower than that of the other models. 
Fig. \ref{fig:inference_times} compares the inference times of the models and CS upon reconstructing a single slice, for up to $\mathit{t}$=16 number of time-steps (Fig. \ref{fig:inference_times_time_steps}) and $F$=256 number of features (Fig. \ref{fig:inference_times_features}). The figure shows that the RIM models performed two to three times faster than CS. For $F$=256 the inference time increases significantly. The IRIM achieves faster inference times than the GRIM and the MRIM for all test cases, because of lower number of parameters. 

Reported SNR values for the training datasets were SNR=24 $\pm$ 6 for the $T_1$-weighted brain dataset, SNR=20 $\pm$ 6 for the $T_{2}^*$-weighted brain dataset, SNR=16 $\pm$ 6 for the $T_2$-weighted knee dataset, and SNR=14 $\pm$ 3 for the independently evaluated $T_2$-weighted TSE brain dataset.

Fig. \ref{fig:SSIM_PSNR_All_Models_10x} depicts box-plots of the SSIM- and PSNR-scores for evaluating performance on different test sets (indicated in the graph headings). Each model was trained using the two loss functions (subsection \ref{subsec:loss_function}) and different dataset compositions (subsection \ref{subsec:Datasets}), as referred to at the top of the figure. 
For $T_1$- and TSE brain data, the $\ell_1$-norm resulted in a slight improvement in performance compared to the $\ell_2$-norm. 
The $T_1$-dataset appears to yield trained networks that generalize well to the other datasets. 
The network trained on $T_{2}^*$-data does not perform well on the $T_{2}$-data but it does show acceptable performance on the other datasets.
Training on $T_2$ knee data results in clearly suboptimal results for all other modalities.
Training on combined (ALL) datasets shows some generalization capabilities but does not yield the best performance.
Overall performance of the models shows a preference for using the $\ell_1$-norm as loss function over the $\ell_2$-norm. 
CS approach generally performs worse than the RIM for all evaluation data and metrics, except for TSE-brains, where it shows a similar SSIM (but lower PSNR).

\begin{figure*}[]
    \centering
    \centerline{\includegraphics[width=\textwidth]{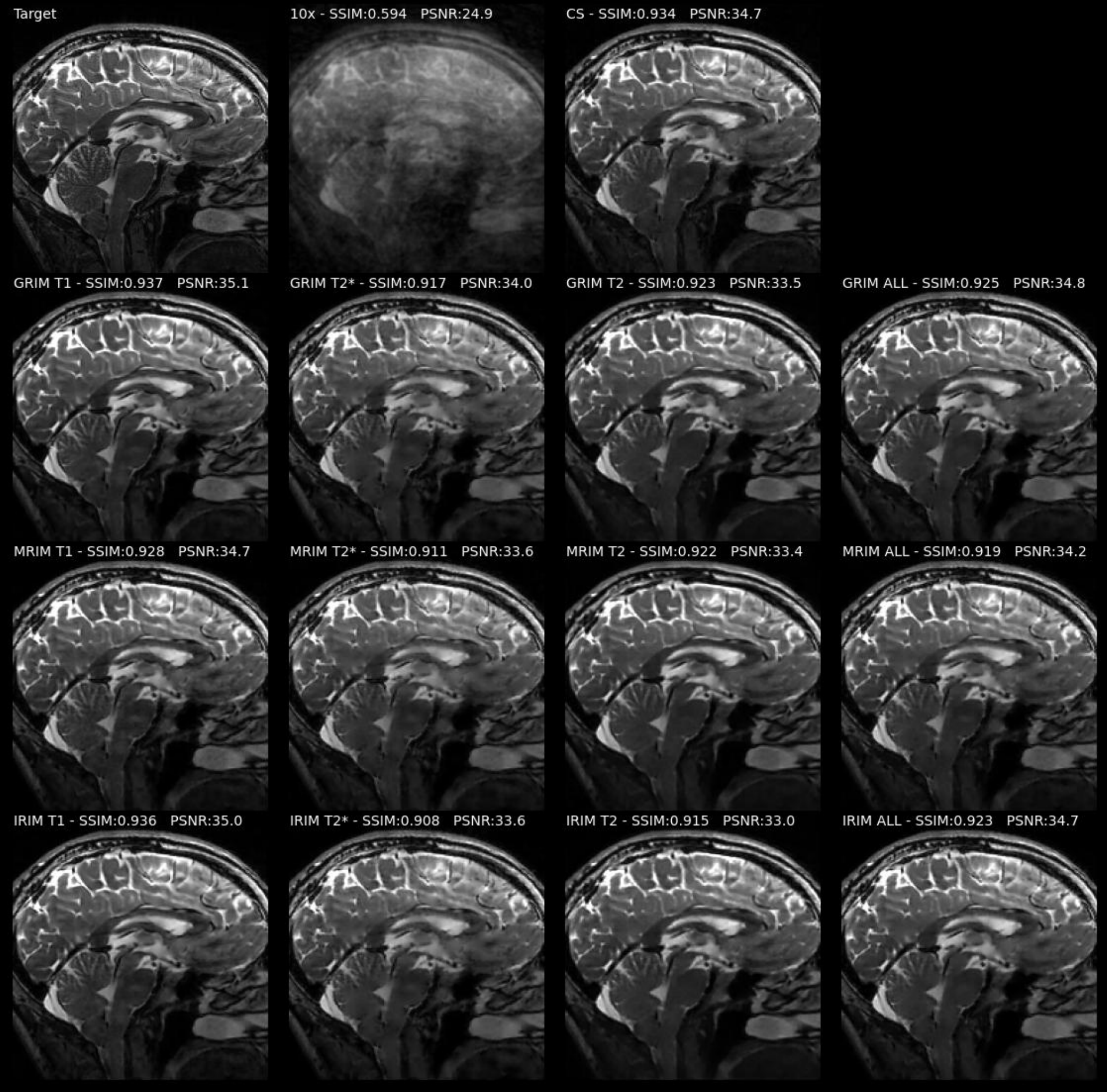}}
    \caption{Reconstructions of the reference-target image (i.e. a fully sampled $T_2$-weighted TSE brain image, first row-first image) by each model applying acceleration factor of 10x. Each model starts the estimation from a direct, aliased reconstruction of the undersampled k-space data (first row, second image). The third image of the first row shows the CS reconstruction. Second, third, and fourth rows show the reconstructions of the GRIM, the MRIM, and the IRIM respectively. From left to right columns show the reconstructions of each model trained with the $\ell_1$-norm, on the $T_1$-weighted brain data , the $T_{2}^*$-weighted brain data, the $T_2$-weighted knee data, and on all datasets, respectively.
    }
    \label{fig:target_&_recons}
\end{figure*}

\begin{figure*}[hbt!]
    \centering
    \centerline{\includegraphics[width=\textwidth]{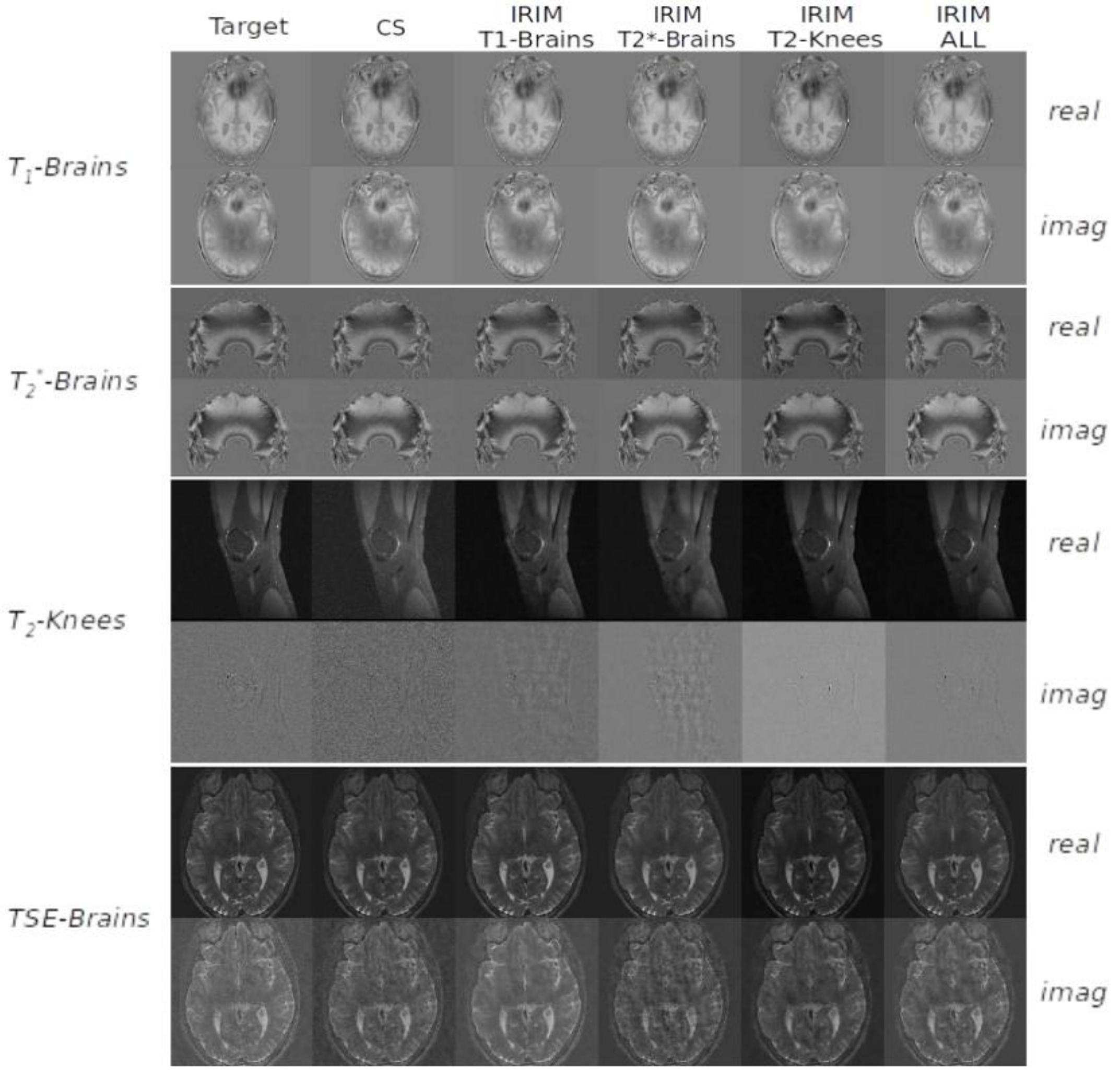}}
    \caption{Reconstructions of the real and imaginary part of the reference-target image in the left-most column, using CS in the second column, and the IRIM trained on the $T_1$-Brains data, $T_{2}^{*}$-Brains data, $T_2$-Knees data, and ALL datasets on third, fourth, fifth, and sixth column respectively. Along the rows are the different types of datasets, each of which is represented through the real and imaginary image parts.}
    \label{fig:real_imag_part}
\end{figure*}

Fig. \ref{fig:target_&_recons} shows the reconstruction images of one slice from the $T_2$-weighted TSE brain dataset by each model trained with the $\ell_1$-norm as loss function using the different training sets. Observe that the less complex MRIM and IRIM models did not yield degraded image quality compared to the GRIM models. Models trained on the $T_{2}^*$-weighted brain dataset appeared to be affected by residual aliasing noise, resulting in inferior image quality compared to the other models.
Example reconstruction of real and imaginary channels of all models and modalities can be found in Suppl. Fig. \ref{fig:real_imag_part}.

\ref{table:grim_mrim_irim_ssim_psnr_scores} summarizes the performances of CS and all models trained with the $\ell_1$-norm, on the TSE evaluation set. Apparently, the models trained on the $T_1$-weighted brain data have the highest SSIM scores and are mostly comparable. CS cannot achieve as high PSNR as the RIM models, with any training data.

\subsection{Robustness}
\begin{figure}[]
    \centering
    \captionsetup[subfigure]{justification=centering}
    \begin{subfigure}[t]{\columnwidth}
        \centering
        \centerline{\includegraphics[width=0.8\columnwidth]{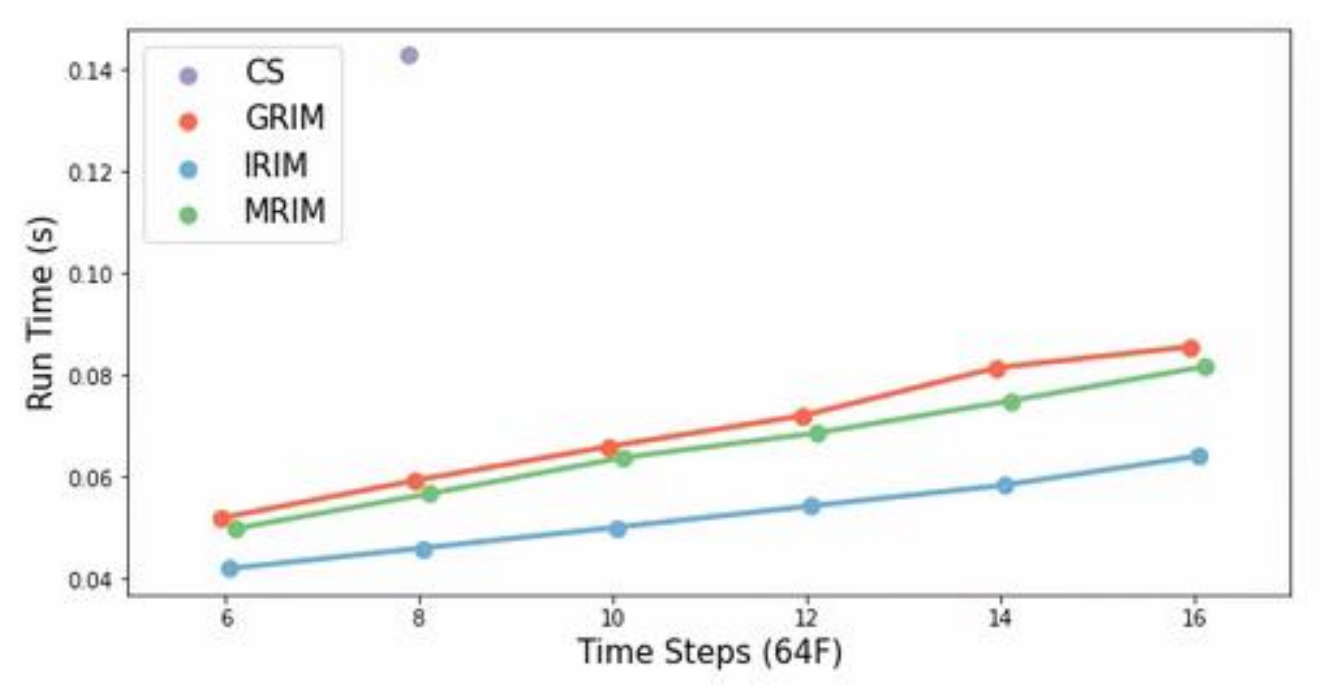}}
        \caption{Inference times as a function of the number of time-steps}
        \label{fig:inference_times_time_steps}
    \end{subfigure}
    \begin{subfigure}[t]{\columnwidth}
        \centering
        \centerline{\includegraphics[width=0.8\columnwidth]{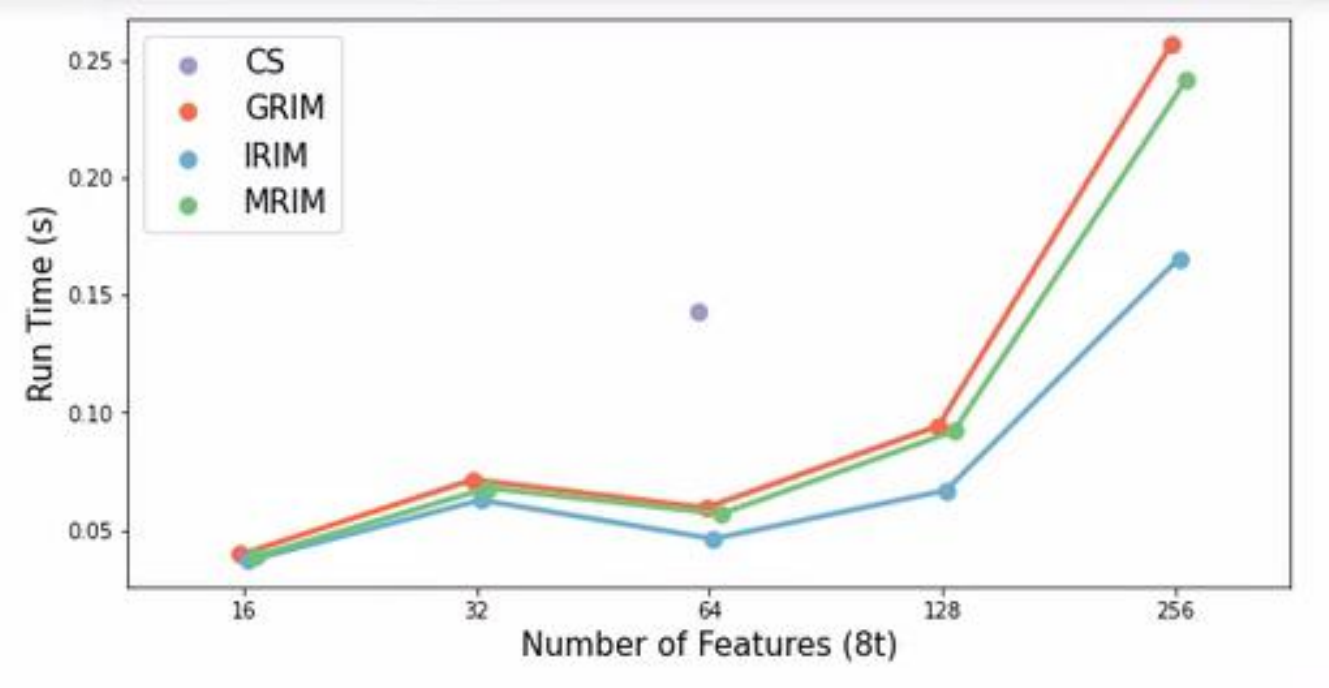}}
        \caption{Inference times as a function of the number of features}
        \label{fig:inference_times_features}
    \end{subfigure}
    \caption{Inference time of CS against the inference times of the three models, (a) as a function of the number of time-steps for 64 features and (b) as a function of the number of features for 8 time-steps. The plots show the mean inference times (\SI{}{\sec}), over 300 repetitions, of each model upon reconstructing a single slice. Standard deviations were negligible, rendering error bars invisible. 
    }
    \label{fig:inference_times}     
\end{figure}

\begin{figure*}[]
    \centering
    \captionsetup[subfigure]{justification=centering}
    \begin{subfigure}{\linewidth}
        \centerline{\includegraphics[width=0.9\linewidth]{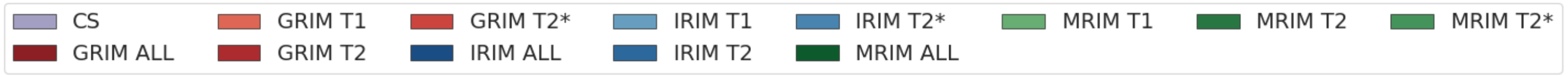}}
    \end{subfigure}
    \\
    \begin{subfigure}[b]{0.45\textwidth}
        \centering
        \centerline{\includegraphics[width=\textwidth]{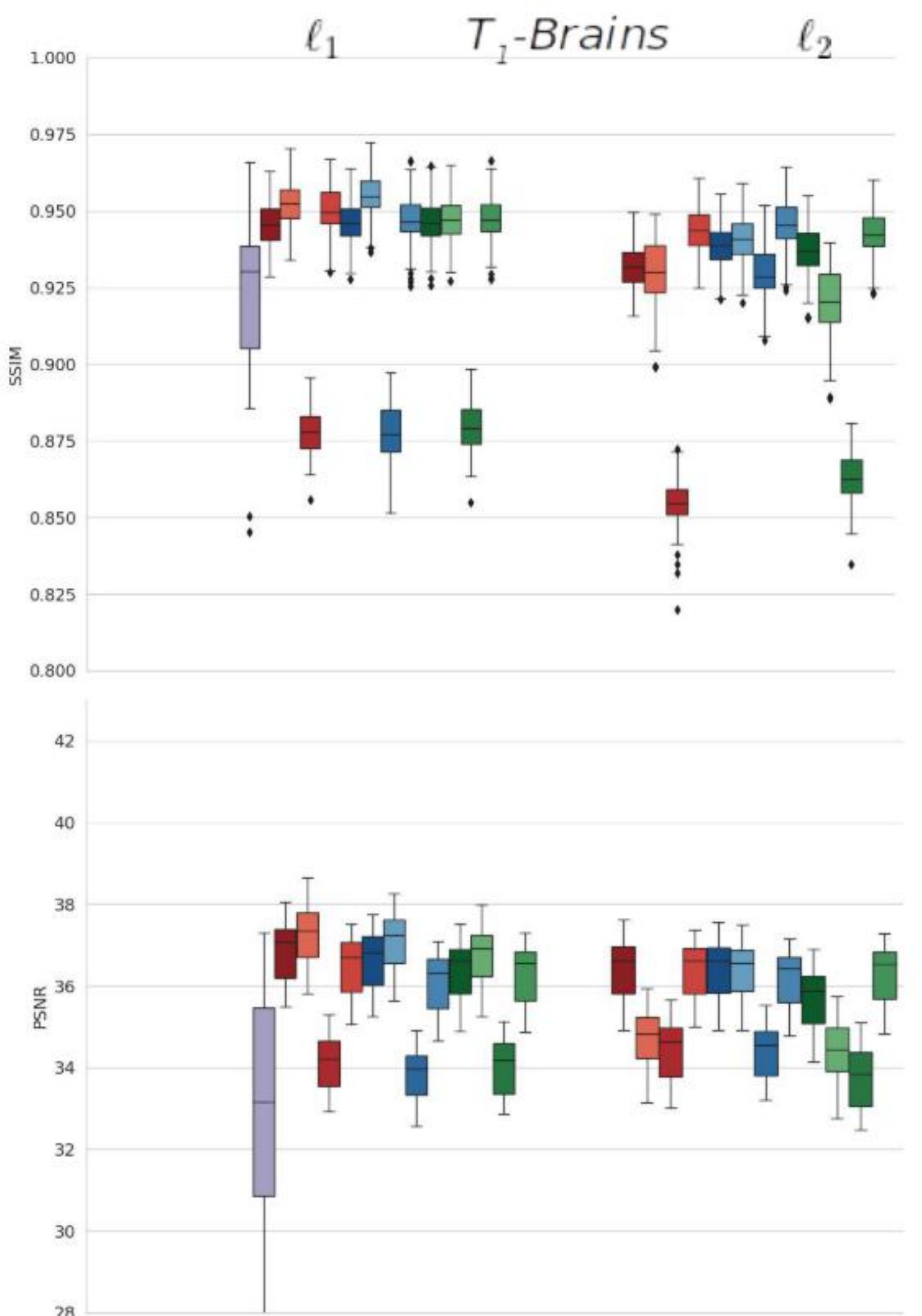}}
        \caption{}
        \label{fig:SSIM_PSNR_All_Models_T1_10x}
    \end{subfigure}
    \hfill
    \begin{subfigure}[b]{0.45\textwidth}
        \centering
        \centerline{\includegraphics[width=\textwidth]{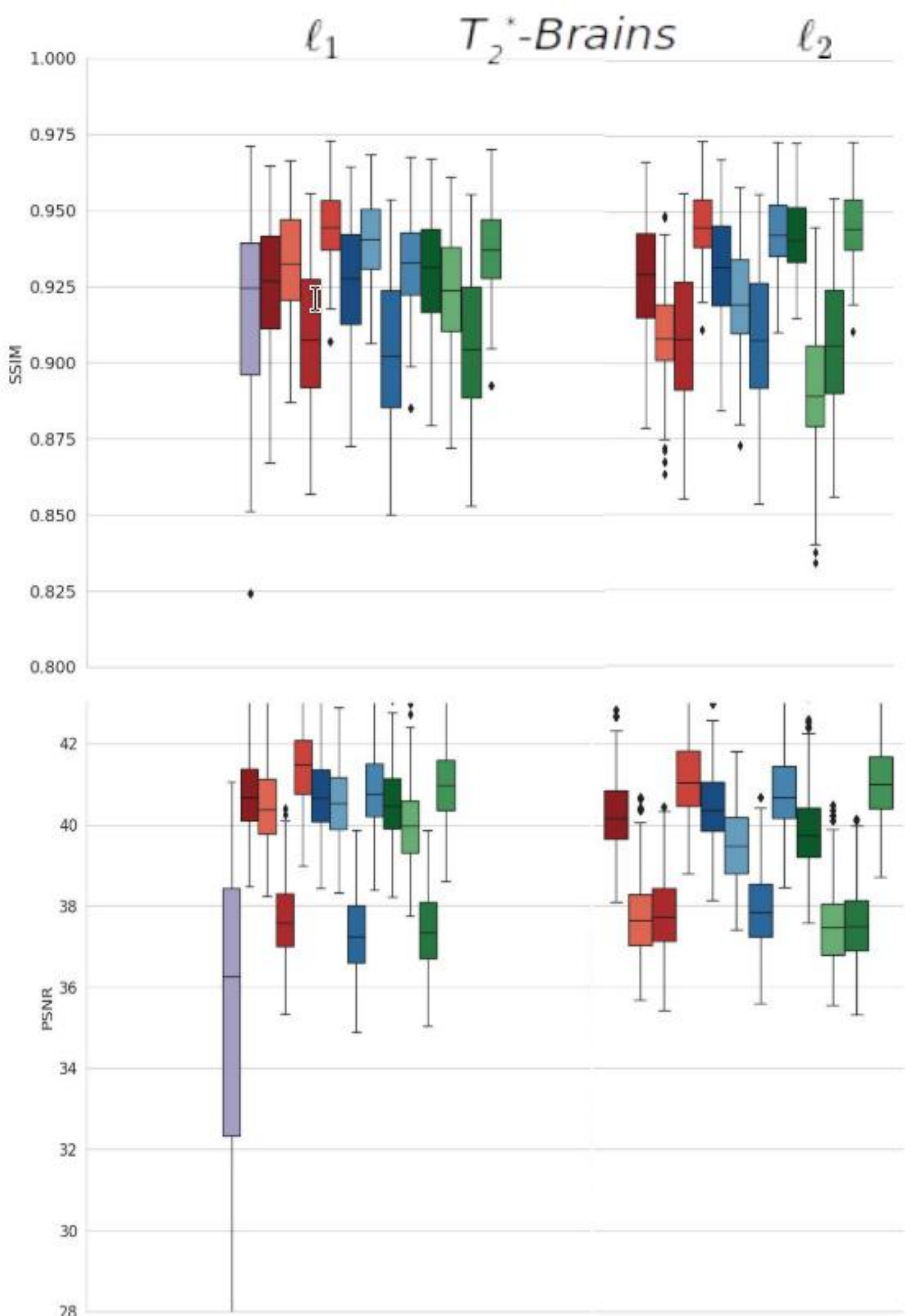}}
        \caption{}
        \label{fig:SSIM_PSNR_All_Models_T2star_10x}
    \end{subfigure}
    \\
    \begin{subfigure}[b]{0.45\textwidth}
        \centering
        \centerline{\includegraphics[width=\textwidth]{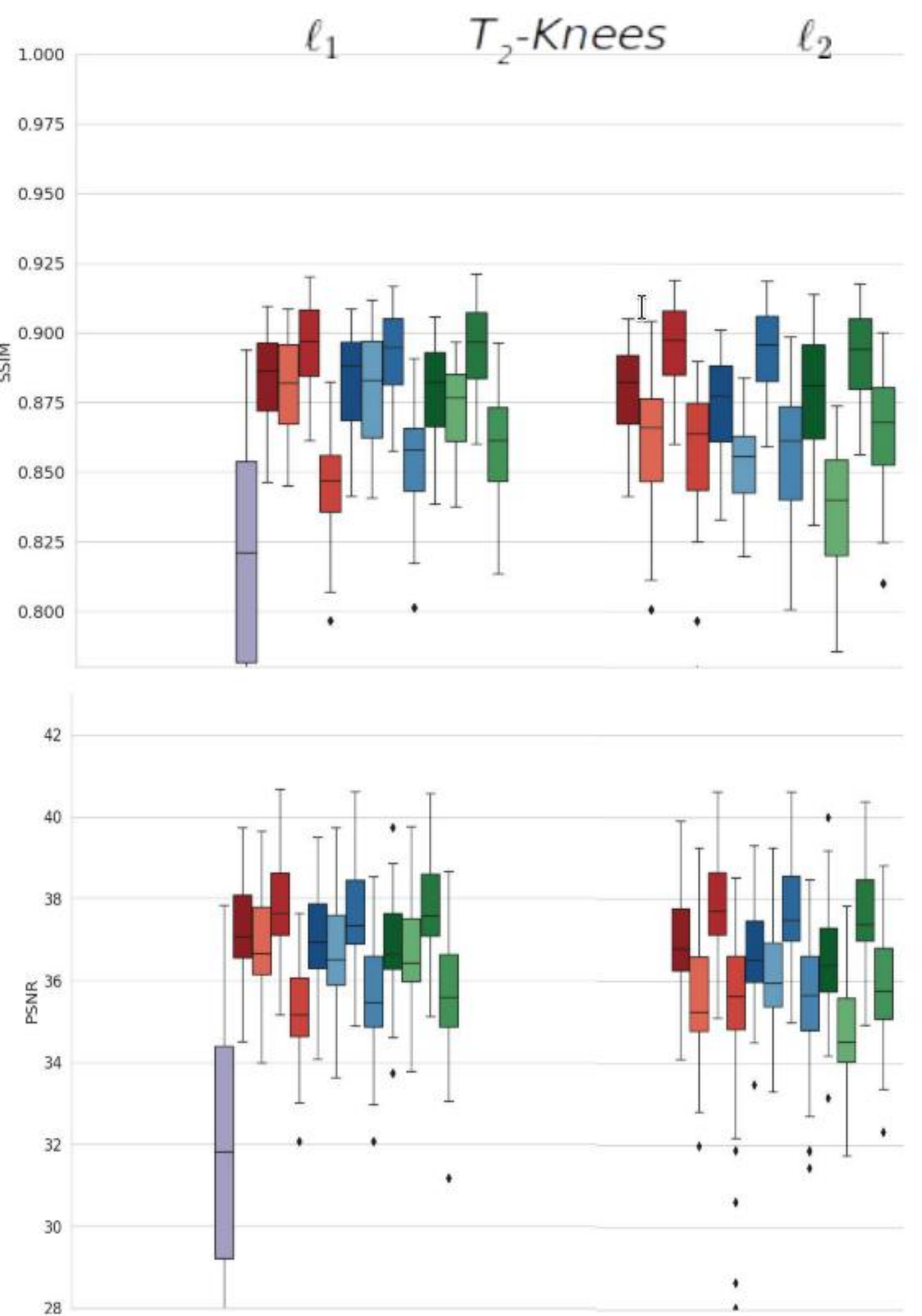}}
        \caption{}
        \label{fig:SSIM_PSNR_All_Models_T2_10x}
    \end{subfigure}
    \hfill
    \begin{subfigure}[b]{0.45\textwidth}
        \centering
        \centerline{\includegraphics[width=\textwidth]{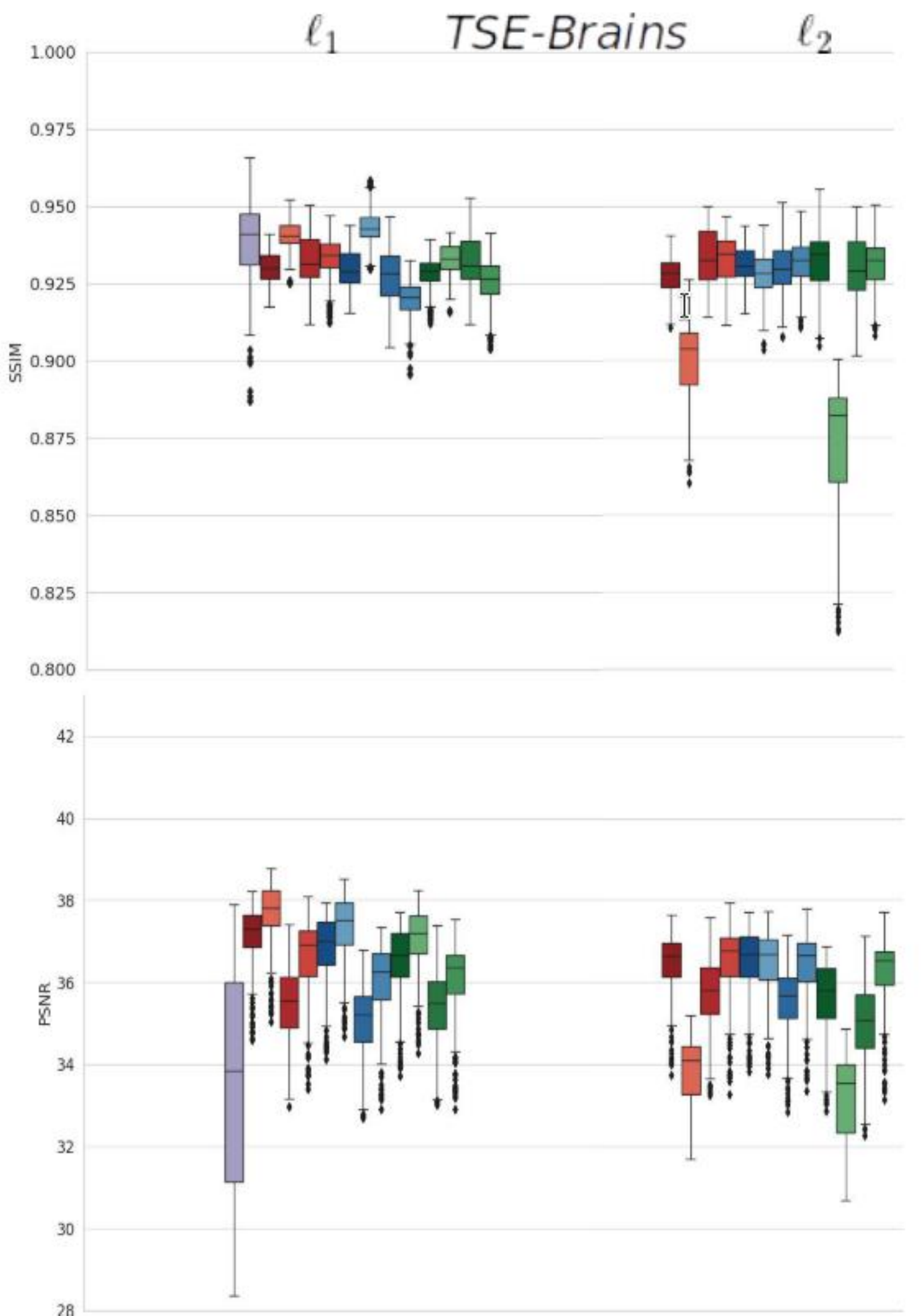}}
        \caption{}
        \label{fig:SSIM_PSNR_All_Models_TSE_10x}
    \end{subfigure}
    \caption{Comparison regarding SSIM and PSNR of CS image reconconstruction against the GRIM, the MRIM and the IRIM networks trained on the three dataset separately and on all three datasets simultaneously (see color legend at the top of the figure). Models were evaluated on (a) the $T_1$-weighted brain data, (b) $T_{2}^*$-weighted brain data, (c) the $T_2$-weighted knee data, and
    (d) the $T_2$-weighted TSE brain data that was never included in training. 
    }
    \label{fig:SSIM_PSNR_All_Models_10x}
\end{figure*}

The most efficient RIM model was selected to test its robustness reconstructing white matter lesions: the IRIM trained with the $\ell_1$-norm as loss function.
This model can perform equivalently or better than the most complex GRIM model (Fig. \ref{fig:SSIM_PSNR_All_Models_10x}), but simultaneously achieved the fastest inference times (Fig. \ref{fig:inference_times}).

\begin{figure*}[]
    \begin{subfigure}{1\textwidth}
        \centering
        \centerline{\includegraphics[width=0.8\textwidth]{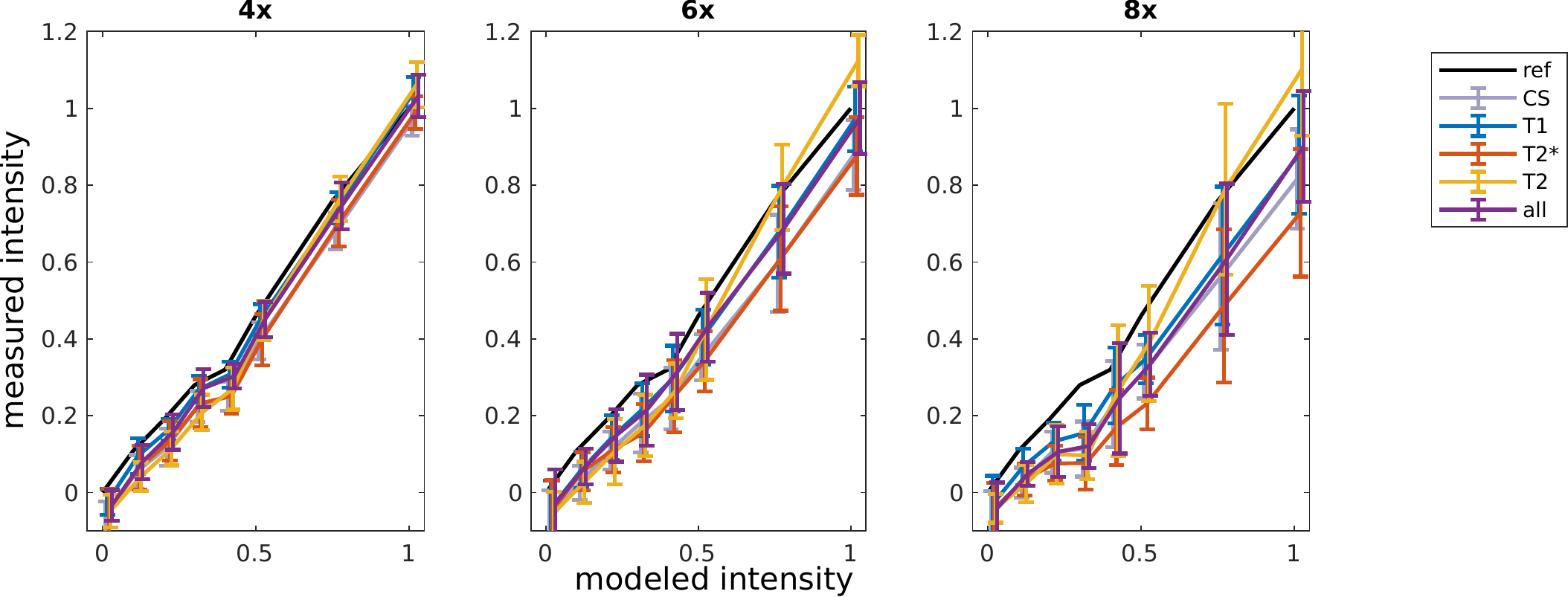}}
        \caption{}
        \label{fig:les_simul_intensity}
    \end{subfigure}
    \begin{subfigure}{0.8\textwidth}
        \centering
        \centerline{\includegraphics[width=\textwidth]{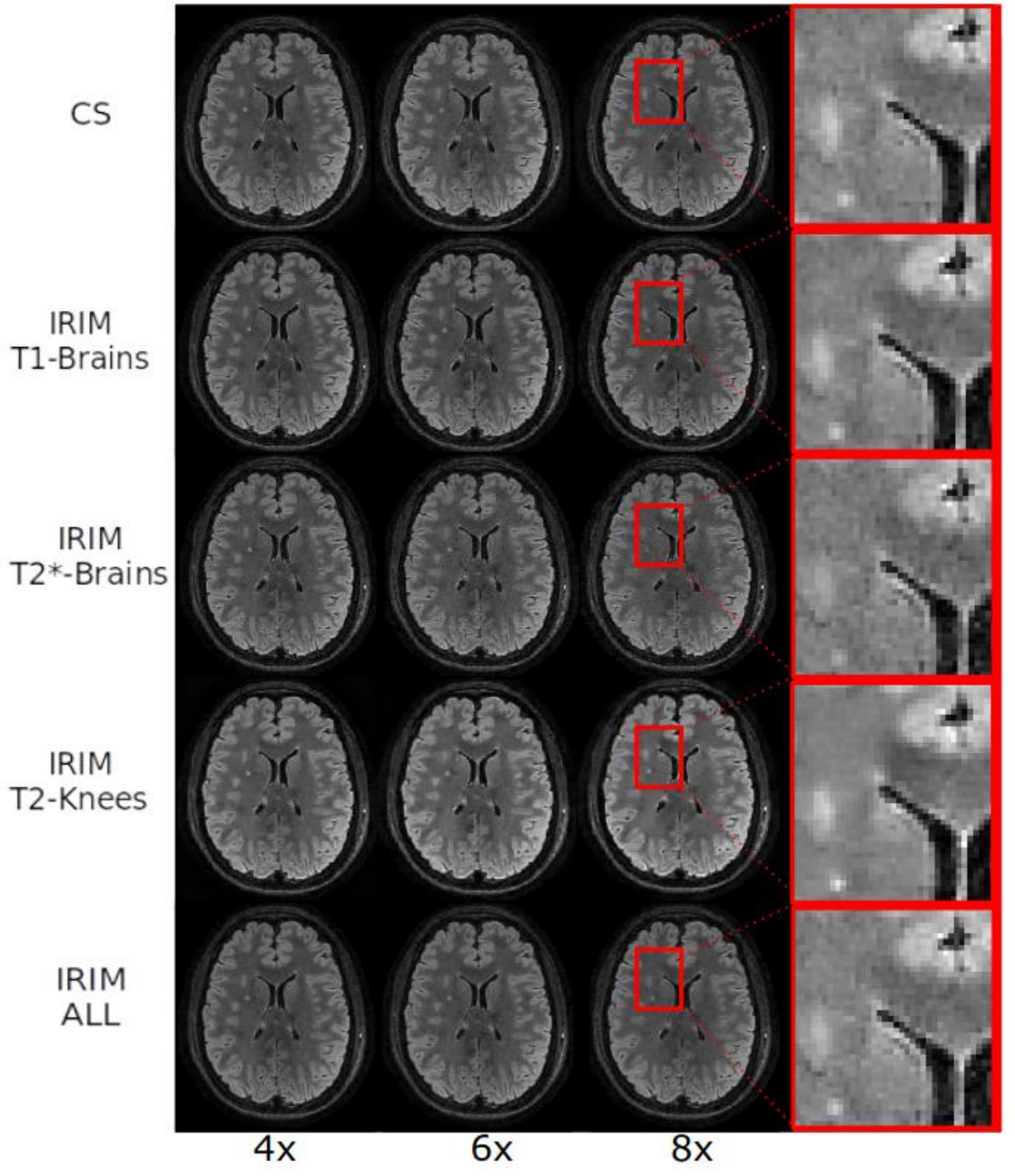}}
        \caption{}
        \label{fig:les_simul_recons}
    \end{subfigure}
    \caption{Simulated lesions reconstructions performed with CS and the IRIM models trained on each dataset composition (subsection \ref{subsec:Datasets}), for 4x, 6x and 8x acceleration factor. (a) Measured lesion intensity as a function of simulated lesion intensity expressed as the multiplication factor of the added Gaussian profile. Error bars denote standard deviations over reconstructions applying 10 random subsampling masks. (b) Reconstructions for lesion intensity of 1.0. The IRIM trained on the $T_2$-Knees dataset shows enhanced contrast and more conspicuous reconstruction of the lesion. Training on the $T_1$-Brains and ALL datasets can perform visually equal to CS, whereas training on the $T_{2}^{*}$-Brains results in a more noisy reconstruction.}
    \label{fig:les_simul}
\end{figure*}

Fig. \ref{fig:les_simul} depicts the results of the lesion simulation experiment. The graphs in Fig. \ref{fig:les_simul_intensity} show the reconstructed intensity in the centre of the lesion as a function of the simulated intensity. Furthermore, different acceleration factors were applied, whereas applying the IRIM trained on the different modalities individually and all of them simultaneously, as well as the conventional CS reconstruction algorithm. Observe that the intensities are expressed through the multiplication factor of the Gaussian profile added to the original signal. Also, noise was added to the data, so that reference intensity fluctuates somewhat, as indicated by the black line. 
For 4-fold acceleration, all models had comparable performance, i.e. within a margin of 5\%, and showed no bias. For higher accelerations the IRIM trained on the $T_2$-weighted knee dataset resulted in the highest intensities, even overestimating the reconstructed lesion intensity by 10\%, whereas CS underestimated it by 40\%. IRIMs trained on the $T_1$-weighted brain dataset and all datasets underestimated the intensity by 15\% and the IRIM trained on the $T_{2}^*$-weighted brain dataset by 45\%.
This also can be observed in the reconstructions in Fig. \ref{fig:les_simul_recons} of the simulated lesion intensity equal to 1.0. The IRIM trained on the $T_1$-weighted brain dataset and ALL datasets show slightly enhanced the signal compared to CS. The IRIM trained on the $T_{2}^*$-weighted brain data yields degraded quality whereas the IRIM trained on the $T_2$-weighted knee dataset is able to reconstruct a brighter lesion with enhanced contrast, but also resulting in a smoother image. 

\begin{figure*}[]
    \centering
    \centerline{\includegraphics[width=\textwidth]{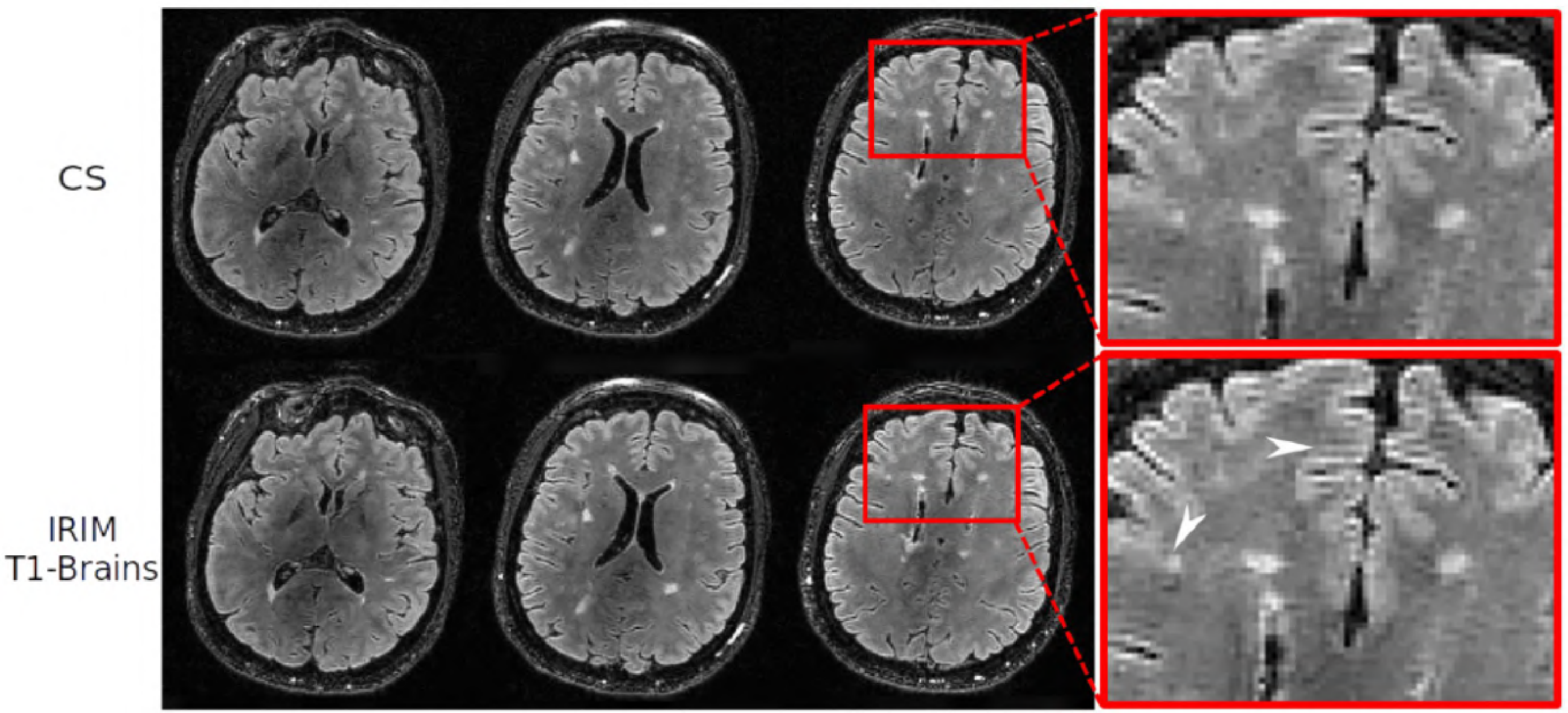}}
    \caption{Prospectively undersampled 3D-FLAIR image of a Multiple Sclerosis patient reconstructed with CS and the most efficient model, the IRIM trained on the $T_1$-Brains dataset with the $\ell_1$-norm, at three axial slices. Arrowheads point to slightly enhanced contrast in the IRIM reconstruction.}
    \label{fig:les_ms}
\end{figure*}

Finally, Fig. \ref{fig:les_ms} shows FLAIR images of an MS patient, reconstructed with the CS and the most efficient model, the IRIM trained on the $T_1$-weighted brain dataset.
Visually, white matter lesions seen in the CS images (here taken as the reference) can also be identified in the IRIM reconstructions. Confirming the simulation experiment, the IRIM slightly enhanced the signal compared to CS. This is visible both in the lesion as well as in the minute cortical gyrus texture amplification (see arrows).

\begin{table}[]
    \centering
    \caption{Total number of parameter of each model as a function of the number of features.}
    \begin{tabular}{@{}
        >{}c 
        >{}l 
        >{}c 
        >{}c 
        >{}c 
        >{}c 
        >{}c @{}}
        \cmidrule(l){1-7}
        \multirow{2}{*}{Model} & \multicolumn{1}{l}{} & \multicolumn{5}{c}{Number of Features} \\ \cmidrule(l){3-7} 
        &  & 256 & 128 & 64 & 32 & 16 \\ \cmidrule(l){1-7}
        GRIM & &1.41M &360k &94.1k &25.2k &7.40k \\
        MRIM &  &1.15M &294k &77.6k &21.4k &6.34k \\
        IRIM &  &753k &196k &53.0k &15.2k  &4.80k \\ 
        \bottomrule
    \end{tabular}    
    \label{table:rim_mrim_indrnn_parameters}
\end{table}

\begin{table*}[]
    \centering
    \caption{Mean SSIM \& PSNR scores with associated standard deviation evaluation  for acceleration factor 10x of CS and the RIM models trained with the $\ell_1$-norm on the unseen, $T_2$-weighted TSE brain data. Each model was trained on the three types of dataset ($T_1$-weighted brain, $T_{2}^*$-weighted brain, $T_2$-weighted knee) separately and on ALL combined. Best scores are indicated with boldface}
    \scalebox{0.8}{
        \begin{tabular}{@{}
            >{}c l
            >{}c 
            >{}c 
            >{}c 
            >{}c 
            >{}c 
            >{}c 
            >{}c 
            >{}c @{}}
            \cmidrule(l){1-10}
             & & \multicolumn{8}{c}{Training Data} \\ \cmidrule(l){3-10} 
             & \multicolumn{1}{l}{} & \multicolumn{2}{c}{$T_1$ brain} & \multicolumn{2}{c}{$T_2^*$ brain} & \multicolumn{2}{c}{$T_2$ knee} & \multicolumn{2}{c}{ALL} \\ \cmidrule(l){3-10} 
            \multirow{-3}{*}{Model} & &\multicolumn{1}{c}{SSIM} & \multicolumn{1}{c|}{PSNR} & SSIM & \multicolumn{1}{c|}{PSNR} & SSIM & \multicolumn{1}{c|}{PSNR} & SSIM & PSNR \\
            \cmidrule(r){1-10}
            GRIM &  &\multicolumn{1}{c}{0.941$\pm$0.01} &\multicolumn{1}{c|}{\textbf{37.6$\pm$1.58}} &0.933$\pm$0.02 &\multicolumn{1}{c|}{36.6$\pm$2.07} &0.932$\pm$0.02 &\multicolumn{1}{c|}{35.4$\pm$1.94} &0.930$\pm$0.01 &37.1$\pm$1.59 \\
            MRIM & &\multicolumn{1}{c}{0.933$\pm$0.01} &\multicolumn{1}{c|}{37.0$\pm$1.65} &0.925$\pm$0.02 &\multicolumn{1}{c|}{36.1$\pm$2.02} &0.932$\pm$0.02 &\multicolumn{1}{c|}{35.4$\pm$1.95} &0.928$\pm$0.01 &36.5$\pm$1.78 \\
            IRIM &  &\multicolumn{1}{c}{\textbf{0.943$\pm$0.01}} &\multicolumn{1}{c|}{{37.3$\pm$1.68}} &0.920$\pm$0.01 &\multicolumn{1}{c|}{36.0$\pm$1.95} &0.927$\pm$0.02 &\multicolumn{1}{c|}{35.1$\pm$1.88} &0.930$\pm$0.01 &36.8$\pm$1.72 \\ 
            \\
            \cmidrule(l){1-10}
            CS &  
            & & & 
            &\multicolumn{1}{c}{SSIM: 0.938$\pm$0.03} &\multicolumn{1}{c}{PSNR: 33.6$\pm$5.46}
            & & & 
            \\ 
            \bottomrule
        \end{tabular}
    }
    \label{table:grim_mrim_irim_ssim_psnr_scores}
\end{table*}

\section{Discussion and Conclusions}
In this paper, we carefully evaluated the efficiency and the dependence on training data of the Recurrent Inference Machine for reconstructing sparsely sampled MRI data, as well as unseen modalities and pathology.
To this end, we presented and compared three architectures with decreasing complexity for the recurrent layers of the model: the Gated Recurrent Unit (GRU), the Minimal Gated Unit (MGU), and the Independently Recurrent Neural Network (IndRNN). Furthermore, we compared the $\ell_1$-norm and the $\ell_2$-norm for the loss function, aiming to make our network learn efficiently and enhance image sharpness in the presence of noise.
We experimented with reconstructions of brain and knee images containing variable contrasts. It was demonstrated that the low-complex Independent Recurrent Inference Machine trained using the $\ell_1$-norm loss-function can accurately reconstruct white matter lesions unseen during training.

The time of reconstructing one slice with a conventional CS algorithm was 144 ms and by using the most efficient Independent Recurrent Inference Machine model based on 8 time-steps and 64 features it was reduced by 68\% to 46.2 ms (see Fig. \ref{fig:inference_times}). Importantly, the IndRNN appeared the most efficient recurrent unit because it involved merely 53k parameters, which meant a reduction on the size of the network by 56\% comparing to the GRU. Simultaneously the reconstructed image quality was well maintained (see Fig. \ref{fig:target_&_recons}).
As such the network design could be particularly useful in time-critical applications, such as in dynamic imaging, during which real-time reconstruction is especially required.

Networks performed best when tested on the data which they were trained on.
Training on the $T_2$-weighted knee dataset was not advantageous when aiming to generalize (Fig. \ref{fig:SSIM_PSNR_All_Models_10x}). However, the knee data contained  homogeneous regions, e.g. muscular tissue as well as sufficient texture or anatomical detail, which seems important to reconstruct structures such as lesions (Fig. \ref{fig:les_simul}) that appear on MS patient data (Fig. \ref{fig:les_ms}).
In contrast, the $T_1$-weighted MPRAGE brain data contained a wide range of textures which might enhance learning the reconstruction, introducing it as stronger training set with generalization capabilities. As such in the more sparsely sampled periphery of k-space, the learned properties aid more in reconstructing high-frequency details of the image.
Training in combined distributions did not result in superior performance, which is in line with earlier work \cite{doi:10.1002/mrm.27355}. Furthermore, the deviation in SNR between the training and the evaluation data led to noise amplification and degradation of the reconstruction.

Importantly, our results demonstrate that the RIM is able to reconstruct images unseen during training. This can be concluded from the results on the $T_2$-weighted TSE brain data that have inherently different contrast. Moreover, it is signified by the RIM's ability to reconstruct FLAIR scans containing lesions of which examples were not present in the training set.
These results confirm that the RIM has learned to solve the inverse reconstruction problem, making it to a large extent invariant to varying image contrasts at inference time. 

Compressed Sensing scored overall lower than the RIM in reconstruction, illustrating that it is advantageous to learn a prior with an efficient model.
In a simulation experiment, the RIM outperformed CS by having a lower bias in lesion intensity. This indicates that it can be advantageous to learn a sparsifying transform from the data, rather than applying a predefined transform as applied in the CS method. This experiment also showed the network's capability to accurately reconstruct small details from undersampled k-space data. 
The experiment with the MS patient data demonstrated that deep learning based image reconstruction is indeed feasible with prospectively acquired, undersampled data. Clearly, a next step would be to further validate the method in a clinical study.

Our work extends other work into reconstruction of MR images using DL methodology, see e.g. \cite{DBLP:journals/corr/abs-1805-10704}.
Such approaches were shown to preserve the natural appearance of the images as a whole whereas also improving depiction of the pathologies compared to reference methods \cite{doi:10.1002/mrm.26977, Geremia2011}. In \cite{10.1007/978-3-319-52280-7_8} the experiments were based on the MICCAI 2009 LV Segmentation Challenge dataset, which consisted of 45 CINE MRI images harboring a range of pathologies. 
Other work exploited uncertainty measures in deep networks for multiple sclerosis lesion detection and segmentation \cite{DBLP:journals/corr/abs-1808-01200}.
We thus add evidence to the claim that there is merit in the clinical use of deep learning for image reconstruction.

Crucially, we aimed to achieve faster inference times and maintain performance by introducing a recurrent unit with lower complexity than those used thus far.
To understand this outcome, we now argue why using the $\ell_1$- instead of the $\ell_2$-norm is needed in training networks with reduced complexity in order to maintain performance. Recent work has shown that the degrees of freedom increase with larger sizes of the training data, making an unrolled network converge slower when trained on big datasets than on fewer training examples \cite{DBLP:journals/corr/DiamondSHW17, DBLP:journals/corr/abs-1906-03742}. Moreover, finding a global minimum with gradient descent is possible for a deep neural network with residual connections (ResNet), despite the objective function being non-convex \cite{DBLP:journals/corr/abs-1811-03804}. 
Likewise, we believe that similar behavior can be observed in a complex unit, such as the GRU. However, for less complex networks, such as the MGU and the IndRNN, the $\ell_2$-norm might slow down the optimization, whereas the $\ell_1$-norm can guide the network's gradients to the right direction, escaping saddle points \cite{doi:10.1063/1.480097} and learn how to solve a high-dimensional problem efficiently.
This is in line with the observed blurring of images reconstructed using the $\ell_2$-norm with lower SNR. Indeed, when using the $\ell_1$-norm, the reconstruction is more robust because of the lower penalization of outliers \cite{Boyd:2011:DOS:2185815.2185816}.

Limitations of our work include the absence of exhaustive testing on other anatomies and modalities than those presented. Also, we did not study variations of the sampling strategy since this was not the aim of this work. Specifically, this has been assessed elsewhere \cite{doi:10.1002/mrm.27355}.
Notably, we have shown before that the RIM can be successfully trained and applied to a range of acceleration factors. Implicitly, we also demonstrated that the RIM generalized to deviating matrix sizes \cite{LONNING201964}.
With the rapid growth of the field, raw data from multiple modalities of the entire human body can be expected to become available shortly. This will allow us to study the generalization properties already currently observed in further detail. Recently, the fastMRI challenge released a large-scale collection of both raw MR measurements and clinical MR images of knees and brains \cite{doi:10.1002/mrm.28338}. Furthermore, the Multi-channel MR Image Reconstruction Challenge made public a large multi-vendor, multi-field-strength brain MR dataset \cite{SOUZA2018482}.

In conclusion, we found that the RIM learns image reconstruction well, making accelerated-MRI feasible and applicable to data, and particular pathologies, unseen during training.
What is more, it offers fast inference times through an efficient and robust approach, thus bringing it closer to the clinical workflow.

\section*{Acknowledgment}
The authors thank B. van Hoek for his support.

\section*{Conflict of interest}
M.W.A. Caan is shareholder of Nico-lab B.V. H.E. Hulst serves on the editorial board of MSJ and has received compensation for consulting services or speaker honoraria from Sanofi Genzyme, Merck Serono, Biogen Idec and Celgene.

\beginsupplement

\end{document}